\documentclass[12pt,a4paper]{article}
\pdfoutput=1
\usepackage{amsmath,amssymb,amsfonts,cancel}
\usepackage{graphicx}
\usepackage{array,booktabs,colortbl,multirow}
\usepackage{colortbl,colordvi,xcolor,color}
\usepackage{rotating}
\usepackage{cite}

\newcommand{\be}{\begin{equation}} 
\newcommand{\ee}{\end{equation}} 
\newcommand{\bea}{\begin{eqnarray}}  
\newcommand{\eea}{\end{eqnarray}}
\newcommand{\bs}{\begin{split}} 
\newcommand{\es}{\end{split}}
\newcommand{\nn}{\nonumber \\} 

\newcommand{\iDslash}{\!~i\!~\cancel{D}\!~}

\newcommand{\lrD}{~\!\overset{\leftrightarrow}{\hspace{-0.1cm}D}\!}
\newcommand{\lrDa}{~\!\overset{\leftrightarrow}{\hspace{-0.1cm}D}\!^{\!~a}}

\newcommand{\U}[1]{U(#1)}
\newcommand{\SU}[1]{SU(#1)}

\newcommand{\units}[1]{~\mathrm{#1}}

\newcommand{\mt}[1]{\mathrm{#1}}

\newcommand{\bfmath}[1]{{\mbox{\boldmath{$#1$}}}}
\newcommand{\Bfmath}[1]{{\large{\mbox{\boldmath{$#1$}}}}}

\newcommand{\vtext}[1]{\begin{sideways}\small{#1}\end{sideways}}

\newcommand{\ctoprule}{\toprule[0.5mm]}
\newcommand{\cbottomrule}{\bottomrule[0.5mm]}
\newcommand{\cmrule}{\midrule[0.25mm]}
\newcommand{\crowcolor}{\rowcolor[rgb]{0.9,0.9,0.9}}

\newcommand{\tcolor}[1]{\color[rgb]{0.4,0.4,0.4}{#1}}

\newcommand{\hc}{\mathrm{h.c.}}

\newcommand{\SM}{\mathrm{SM}}

\newcommand{\BLOp}{$\cancel{\mt{B}}$ \& $\cancel{\mt{L}}$}
\newcommand{\BLOpbf}{$\bfmath{\cancel{\mt{B}}}$\bf{ \& }$\bfmath{\cancel{\mt{L}}}$}

\begin{document}

%----------------------------------- TITLE AND AUTHORS -----------------------------------------%

%Preprint numbers
\begin{flushright}
CERN-PH-TH-2014-264\\
DESY 15-011\\
\today
\end{flushright}
\vspace*{5mm}
\begin{center}

\renewcommand{\thefootnote}{\fnsymbol{footnote}}
\setcounter{footnote}{1}

{\Large {\bf Observable Effects of General New Scalar Particles
}} \\
\vspace*{0.75cm}

{\bf J.\ de Blas${}^a$\footnote{E-mail: Jorge.DeBlasMateo@roma1.infn.it},}
{\bf M.\ Chala${}^b$\footnote{E-mail: mikael.chala@desy.de},}
{\bf M.\ P\'erez-Victoria${}^c$\footnote{E-mail: mpv@ugr.es}} and 
{\bf J.\ Santiago${}^{c,d}$\footnote{E-mail: jsantiago@ugr.es}}
\vspace{0.5cm}

{\it ${}^a$INFN, Sezione di Roma, Piazzale A. Moro 2, I-00185 Rome, Italy}\\
\vspace{0.5cm}
{\it ${}^b$DESY, Notkestrasse 85, 22607 Hamburg, Germany}\\
\vspace{0.5cm}
{\it ${}^c$Departamento de F\'{\i}sica Te\'orica y del Cosmos and CAFPE,\\
Universidad de Granada, Campus de Fuentenueva, E-18071 Granada, Spain}\\
\vspace{0.5cm}
{\it ${}^d$CERN, Theory Division, CH1211 Geneva 23, Switzerland}\\ 
\vspace{0.5cm}

\end{center}
\vspace{0.5cm}

%--------------------------------------------- ABSTRACT ---------------------------------------------%
\begin{abstract}
 
\noindent We classify all possible new scalar particles that can have
renormalizable linear couplings to Standard Model fields and therefore be singly
produced at colliders. We show that this classification exhausts the list
of heavy scalar particles that contribute at the tree level to the
Standard Model effective Lagrangian to dimension six. 
We compute this effective Lagrangian for a general scenario 
with an arbitrary number of new scalar particles
and obtain flavor-preserving constraints on their couplings and masses. 
This completes the tree-level matching of the coefficients of dimension five and six operators
in the effective Lagrangian to arbitrary extensions of the Standard Model.

\end{abstract}

\renewcommand{\thefootnote}{\arabic{footnote}}
\setcounter{footnote}{0}

%-------------------------------- DOCUMENT: INTRODUCTION ---------------------------------%

\clearpage

\section{Introduction}
\label{section_Intro}

The discovery of the Higgs boson at the Large Hadron Collider
(LHC)~\cite{Aad:2012tfa,Chatrchyan:2012ufa} has opened a new era in
particle physics in which 
we have, for the first time, direct access to the electroweak symmetry
breaking sector. The naturalness
problem associated to this 
scalar sector is still one of the main reasons to expect new
physics beyond the Standard Model (SM) at the TeV scale, and thus
accessible to the LHC. 
However, the lack of significant deviations from the SM predictions
after Run 1 suggests that, even if really present, the new particles  
may be too heavy to be produced on-shell, 
so that only their indirect effects can be observed at the
LHC ---although it is certainly possible that they are just above the
current reach 
and can still be directly produced at the higher energies of Run 2. In
such a case, the natural language to parameterize the 
expected effects of new physics is that of effective theories. 

Effective Lagrangians provide a model-independent description of the
effects of new particles at energies much smaller than their masses.
Hence, they are the perfect tool to study any new physics that lies
beyond the reach of our current experiments. 
A single higher-dimensional gauge-invariant operator contributes in
general to several different couplings between SM particles 
after electroweak symmetry
breaking, leading to non-trivial correlations
between different observables~\cite{delAguila:2000aa}. These
correlations can be tested 
experimentally or, alternatively, be used to predict the size of
the expected deviations with respect to the SM
predictions~\cite{Pomarol:2013zra}. In addition, 
higher-dimensional operators are often generated by a smaller
number of couplings in specific ultraviolet completions 
and therefore the coefficients of different
operators are correlated as well.  
It could be argued that these latter correlations are model-dependent, 
defeating the very purpose of the effective
Lagrangian. However, model independence can be recovered 
if a complete dictionary between ultraviolet completions and effective
operators is built. Such a dictionary would provide a
comprehensive classification of new physics with potentially
observable effects at the 
LHC. This classification can hence guide experimental searches to ensure that no
viable option is missed at the LHC, and help
to identify the origin of possible deviations from the SM predictions.

Encouraged by the recent experimental observation of a Higgs sector, in this article we focus on new scalar particles. We first
classify all the possible new scalars that can couple linearly, with renormalizable interactions, to SM
fields, and write their most general phenomenologically-relevant interactions. These particles can be singly produced at colliders with sizable couplings and have therefore the most promising discovery potential at the LHC. 
The corresponding general interactions of new quarks, leptons and
vector bosons have already been given in
Refs.~\cite{delAguila:2000rc},~\cite{delAguila:2008pw} and
~\cite{delAguila:2010mx}, respectively. Our results here thus complete
the description of arbitrary new particles with linear gauge-invariant
renormalizable couplings to the SM fields. This provides an extremely
useful scheme for a model-independent interpretation of LHC
searches. Particular models correspond to specific choices of the
general couplings and masses in this set-up. Therefore, once the
experimental results are written in terms of the general parameters it
is straightforward to derive consequences for any model of
choice. 

It turns out that the same classification also covers all
possible new scalar particles that contribute, once integrated out at the tree level, to the SM
effective Lagrangian of dimension five and six. We perform this integration explicitly.
In predictive models loop contributions are suppressed, so it is expected that the leading observable
consequences of new heavy scalars are those generated at tree level.
Except for this assumption, our results are completely general and can
be used for an arbitrary extension of the SM with heavy scalars, 
independently of their amount and quantum numbers. Furthermore, extra
heavy particles with different spins do not mix with the heavy scalars
in their contribution to the dimension-six effective Lagrangian at
tree level.
Therefore, together with the effective Lagrangians generated by the most general
extension of the SM with new quarks~\cite{delAguila:2000rc},
leptons~\cite{delAguila:2008pw}
and vector bosons~\cite{delAguila:2010mx}, our results complete 
the tree-level dictionary between any model of new physics
and the dimension-six SM effective Lagrangian.
This dictionary can be used to trivially obtain the observable
implications of an arbitrary model of new physics at energies much smaller than
the masses of the new particles involved. It is also a powerful
tool to investigate correlations or cancellations predicted among
observables in specific extensions of the SM. Finally, it provides a
rationale for the calculation of the constraints on the coefficients of
the SM effective Lagrangian, as one can put constraints on the sources
of the effective operators in a correlated way 
rather than on arbitrary combinations of them. 

The article is organized as
follows. We classify in Section~\ref{section_Extra_Scal} all possible
new scalars that can have linear interactions with SM fields. We
describe in Section~\ref{section_Eff_Lag} how to compute the
effective Lagrangian that results from the tree-level integration of an
arbitrary number and type of new heavy scalars, and show that the previous
classification exhausts the list of new spin-0 particles that can contribute to
the dimension-six effective Lagrangian at tree level. In Section~\ref{section_ObsEff}
we discuss what effects induced by the heavy scalars are observable at this order. 
Some applications of the effective Lagrangian are presented in
Section~\ref{section_Pheno}. Finally, we comment on the interplay between
new scalars and particles with different spin in Section~\ref{section_NSNVNF}, 
and conclude in Section~\ref{section_Conclusions}. The basis of dimension-six operators
and all the relevant scalar interactions with the
resulting effective Lagrangians, are given in Appendix~\ref{app: L6_basis} and 
\ref{app: NS_OpCoeff}, respectively.

%-------------------------------- DOCUMENT: SECTIONS -----------------------------------------%

\section{Standard Model extensions with extra scalar fields}
\label{section_Extra_Scal}

We consider a general renormalizable theory for extra scalars and SM fields, invariant under Lorentz transformations and under the complete $\SU{3}_c\otimes \SU{2}_L \otimes \U{1}_Y$ gauge group. The new scalar fields will come in complete representations of this group, which can be decomposed into their irreducible components $\sigma$. Non-renormalizable interactions are also possible in principle, but in a predictive theory they will be suppressed by a scale larger than the mass of the extra scalars. Here we concentrate on the leading effects, which are generically described by operators of dimension four at most. 

The most general Lagrangian for such an extension of the SM can be written as 
\be
{\cal L}={\cal L}_{\SM}+{\cal L}_\sigma+{\cal L}_{\mathrm{int}},
\label{FullL}
\ee
where ${\cal L}_{\SM}$ is the SM Lagrangian, ${\cal L}_\sigma$ contains the kinetic  (with covariant derivatives) and mass terms for the new scalars and ${\cal L}_{\mathrm{int}}$ describes the non-gauge interactions of the extra scalars.

The SM Lagrangian reads, in standard notation\footnote{We use capital indices $A,B,C$ as color indices, whereas lower case indices $a,b=1,2,3$ tag fields in the adjoint of $SU(2)_L$. Latin indices $i,j,k$ are used to label different generations.}
\be
\begin{split}
{\cal L}_{\SM}=&-\frac 14 G^A_{\mu\nu}G^{A\ \mu\nu}-\frac 14 W^a_{\mu\nu}W^{a\ \mu\nu}-\frac 14 B_{\mu\nu}B^{\mu\nu}\\
+&\overline{l_L^i} \iDslash  l_L^i+\overline{q_L^i} \iDslash
q_L^i+\overline{e_R^i} \iDslash e_R^i+\overline{u_R^i} \iDslash
u_R^i+\overline{d_R^i} \iDslash d_R^i\\ 
+&\left(D_\mu \phi\right)^\dagger D^\mu \phi-U\left(\phi\right)-\left(y^e_{ii}~\overline{l_L^i}\phi e^i_R+y^d_{ii}~\overline{q_L^i}\phi d^i_R+V^\dagger_{ij}y^u_{jj}~\overline{ q_L^i}\tilde\phi u^j_R+\hc\right).
\label{SMLag}
\end{split}
\ee
In order to fix the meaning of flavor indices, we have chosen a basis
in which the Yukawa interactions for the charged-leptons and down-type
quarks are diagonal. As usual,  ${\tilde \phi}=i\sigma_2 \phi^*$ denotes the iso-doublet of hypercharge $-1/2$, constructed with the Higgs doublet $\phi$. 
The Higgs scalar potential is
\be
U\left(\phi\right)=-\mu_\phi^2\left|\phi\right|^2+\lambda_\phi\left|\phi\right|^4.
\label{Scalar potential 1}
\ee
 The Lagrangian ${\cal L}_\sigma$ contains the gauge-invariant kinetic and mass terms for the new scalars:
 \be
 {\cal L}_\sigma=\sum_\sigma \eta_\sigma \left[\left(D_\mu \sigma\right)^\dagger D^\mu \sigma-M_\sigma^2\sigma^\dagger \sigma \right],
 \label{Lvarphi}
\ee
where $\eta_\sigma=1,\frac 12$, for complex and real scalars, respectively.  Note that we are working in a basis with canonical kinetic terms and diagonal mass matrices for all scalar fields, including the Higgs doublet. To match models written in a different basis, the diagonalization must be performed prior to using our formulas.\footnote{Furthermore, we assume that there are no tadpole operators in the electroweak symmetric phase. This entails no loss of generality since these tadpoles, which are only possible for new singlet scalars, can always be eliminated by a shift of the singlet field(s). The only effect of this shift is a redefinition of the parameters that we write explicitly.} 

Finally, ${\cal L}_{\mathrm{int}}$ contains the renormalizable
interactions of the extra scalars (among themselves and with the SM
fields), except for the gauge interactions, which are already included
in ${\cal L}_\sigma$. 
We can distinguish between interactions with fermions and purely scalar interactions:
\be
{\cal L}_{\mathrm{int}}=- V(\{\sigma\}, \phi) - \sum_\sigma \eta_\sigma \left(\sigma^\dagger J_\sigma +\hc \right).
\label{LSMvarphi}
\ee
The chirality-flipping fermionic currents $J_\sigma\sim\overline{\psi_L} \otimes \xi_R$ or $J_\sigma \sim \overline{\xi_R} \otimes \psi_L$ couple to one scalar field with a dimensionless coupling.  The potential $V$ contains scalar interactions between the new particles and, possibly, the SM Higgs fields. Together with the mass terms and $U$, it forms the total scalar potential. As explained above, $V$ does not include mass mixing terms, since we work in a basis with diagonal quadratic terms. Thus, each term in $V$ contains either three or four scalars, with couplings of dimension one or zero, respectively. Furthermore, all the terms in $V$ have at least one $\sigma$ field. 

% New Scalar representations
\begin{table}[tbp]
\begin{center}
{\small
\begin{tabular*}{13.5cm}{ l  c  c  c  c  c  c  c  c} 
\ctoprule
\crowcolor\!\!Colorless\hspace{-0.05cm}~&${\cal S}$&${\cal S}_1$&${\cal S}_2$&$\varphi$&$\Xi_0$&$\Xi_1$&$\Theta_1$&$\Theta_3$\\
\crowcolor\!\!Scalars\!\!\!\!&&&&&&&&\\[-0.05cm]
\cmrule
\!\!Irrep\!\!\!\!&$\left(1,1\right)_0$&$\left(1,1\right)_1$&$\left(1,1\right)_2$&$\left(1,2\right)_{\frac 12}$&$\left(1,3\right)_0$&$\left(1,3\right)_1$&$\left(1,4\right)_{\frac 12}$&$\left(1,4\right)_{\frac 32}$\\
&&&&&&\\[-0.495cm]
\cbottomrule
&&&&&&&\\[-0.25cm]
\end{tabular*}}
\\
\vspace{-0.15cm}
\hspace{-0.01cm}
{\small
\begin{tabular*}{13.5cm}{ l  c  c  c  c  c  c } 
\ctoprule
\crowcolor\!\!Colored~~&~~~~~${\omega}_{1}$~~~~~&~~~~~${\omega}_{2}$~~~~~&~~~~~${\omega}_{4}$~~~~~&~~~~~$\Pi_1$~~~~~&~~~~~$\Pi_7$~~~~~&~~~~$\zeta$~~~~~~\\
\crowcolor\!\!Scalars&&&&&&\\[-0.05cm]
\cmrule
\!\!Irrep&$\left(3,1\right)_{-\frac 13}$&$\left(3,1\right)_{\frac 23}$&$\left(3,1\right)_{-\frac 43}$&$\left(3,2\right)_{\frac 16}$&$\left(3,2\right)_{\frac 76}$&$\left(3,3\right)_{-\frac 13}$\\
&&&&&&\\[-0.495cm]
\cbottomrule
&&&&&\\[-0.36cm]
\ctoprule
\crowcolor\!\!Colored&$\Omega_{1}$&$\Omega_{2}$&$\Omega_{4}$&$\Upsilon$&$\Phi$&$ $\\
\crowcolor\!\!Scalars&&&&&&\\[-0.05cm]
\cmrule
\!\!Irrep&$\left(6,1\right)_{\frac 13}$&$\left(6,1\right)_{-\frac 23}$&$\left(6,1\right)_{\frac 43}$&$\left(6,3\right)_{\frac 13}$&$\left(8,2\right)_{\frac 12}$\\
&&&&&&\\[-0.495cm]
\cbottomrule
\end{tabular*}
}\vspace{-0.1cm}
\caption{Scalar bosons with linear renormalizable interactions with the SM fields. The quantum numbers $(R_c,R_L)_Y$ denote the irreducible representation (Irrep) $R_c$ under $SU(3)_c$, $R_L$ under $SU(2)_L$ and the hypercharge $Y$, respectively. The hypercharge is normalized such that the electric charge is $Q=Y+T_3$. Looking only at the quantum numbers, some readers might miss in this list a scalar particle transforming as a $\left(1,2\right)_{\frac 32}$ and coupling linearly to three Higgs doublets. However, the corresponding operator actually vanishes, since it involves an antisymmetric combination of the Higgs fields.\vspace{-0.69cm}}
\label{table:newscalars}
\end{center}
\end{table}

The new scalars with linear interactions in ${\cal L}_{\rm{int}}$ can
be singly produced at tree level in colliders.\footnote{Other scalar
  fields can also be singly produced if they mix with these after
  electroweak symmetry breaking, but if they are heavy the production
  rate will be suppressed by the square of small mixings.} By ``linear
interactions" we mean that ${\cal L}_{\mathrm{int}}$ contains some
non-vanishing term that is the product of SM fields and a single power
of the given extra scalar field, with no other extra
scalars. Gauge
invariance and the particle content of the SM strongly constraints the
quantum numbers of new scalars that can have such linear
interactions. 
As we show in the next section, only the scalars in representations that allow for these linear couplings can contribute at tree level, to order $1/M_\sigma^2$ in the heavy mass limit, to observable processes with SM particles in the initial and final states. We list in~Table~\ref{table:newscalars} all the irreducible representations of scalars with linear interactions of this kind. 
The scalar representations $\cal S$, $\Xi_0$, $\Theta_1$ and $\Theta_3$ do not couple to the SM fermions.
In Appendix~\ref{app: NS_OpCoeff} we write the interactions in ${\cal
  L}_{\mathrm{int}}$ explicitly, including only those that have an
impact in the SM effective Lagrangian at dimension six. These include
the mentioned linear interactions of the scalar fields in the
irreducible representations of~Table~\ref{table:newscalars}, as well
as terms involving two or three scalar fields in the same set of
representations. Note that, in particular, new scalars can
always couple with the SM through a Higgs-portal type of coupling 
$\left(\sigma^\dagger \sigma\right)\left(\phi^\dagger \phi\right)$ \cite{Silveira:1985rk}.
However, only
the ones in our list with the right quantum numbers to allow for linear
couplings to the SM will induce dimension-six operators when integrated
at tree level.
Moreover, as we explain in the next section, interactions
involving two or more new scalars are relevant for the calculation of the effective Lagrangian
to dimension six only in those cases where the scalar potential contains trilinear couplings of new scalars to two Higgs fields (see Tables \ref{Table:S0Table}, \ref{Table:Xi0Table}, \ref{Table:Xi1Table} and \ref{Table:MultiScalTable} in Appendix~\ref{app: NS_OpCoeff}).

Extensions of the SM with new scalar fields have been
extensively considered in the past and partial classifications have
been presented, with special emphasis on their effect on baryon and
lepton number
violation~\cite{Nieves:1981tv,Buchmuller:1986zs,Ma:1998pi}, collider
physics~\cite{DelNobile:2009st,Han:2010rf}, top
physics~\cite{AguilarSaavedra:2011vw} or flavor
physics~\cite{Grinstein:2011dz}. 
However the complete classification of the
scalar fields that can couple linearly to the SM and the calculation
of the tree-level dimension-six effective Lagrangian for the most
general extension of the SM with an arbitrary number of new scalar
fields has, to the best of our knowledge, never been presented before.

%------------------------------------------------------------------------------------------

\section{The effective Lagrangian for heavy new scalar particles}
\label{section_Eff_Lag}

In this paper we are mainly interested in the effects of {\em heavy} new scalars, with masses $M_\sigma$ that are large in comparison to the Higgs vacuum expectation value (vev) and to the energies probed by the available experimental data. An efficient way to describe these effects at leading order is to integrate the heavy scalars out and expand in $1/M_\sigma$ to obtain an effective Lagrangian with gauge-invariant local operators of dimension up to six. This allows for a direct comparison with model-independent analyses and also for a simple combination with other extensions of the SM. In our theory with general extra scalars, the coefficients of these operators will be simple functions of the couplings and masses of the heavy scalars. Once again, we work at the tree level for simplicity and to avoid further suppressions (see \cite{Skiba:2010xn,Han:2010rf,Henning:2014wua} for examples of one-loop integration of scalar multiplets). 

In this section we describe the integration of the heavy scalar fields. In particular, we show that among the infinite scalar representations that can appear in ${\cal L}_{\mathrm{int}}$, only the ones in Table~\ref{table:newscalars} contribute to the effective Lagrangian to dimension six. For clarity, our discussion will be slightly schematic. The complete explicit results, i.e.\ the dimension-six operators and the values of their coefficients, are collected in the appendices. 

In the following we assume that, before electroweak symmetry breaking, the scalar mass matrix has only one negative eigenvalue, $-\mu_\phi^2$. The eigenvector is a $(1,2)_{1/2}$ scalar field $\phi$, which we identify with the SM scalar doublet. The other eigenvalues, $M_\sigma^2$, are assumed to be large in comparison with the Higgs vev and with the relevant energies. Finally, we assume that the dimensionful couplings $\kappa_\sigma$ that multiply the dimension-three operators in $V$ are at most of the size of the smaller heavy-scalar mass, $|\kappa_\sigma| \lesssim M$, with $M = \mathrm{Min}\, \{|M_\sigma|\}$. These assumptions are well motivated by the agreement, within the available precision, of experimental results and SM predictions for Higgs observables. They lead to a decoupling scenario and allow us to perform the integration in the electroweak symmetric phase, which is extremely convenient. The occurrence of electroweak symmetry breakdown and all its effects are captured to order $1/M^2$ by the effective Lagrangian. This includes the case in which the extra scalars acquire (suppressed) vevs in the Higgs phase~\cite{Skiba:2010xn}.

\begin{figure}[t]
\hspace{-1cm}\includegraphics[scale=0.95]{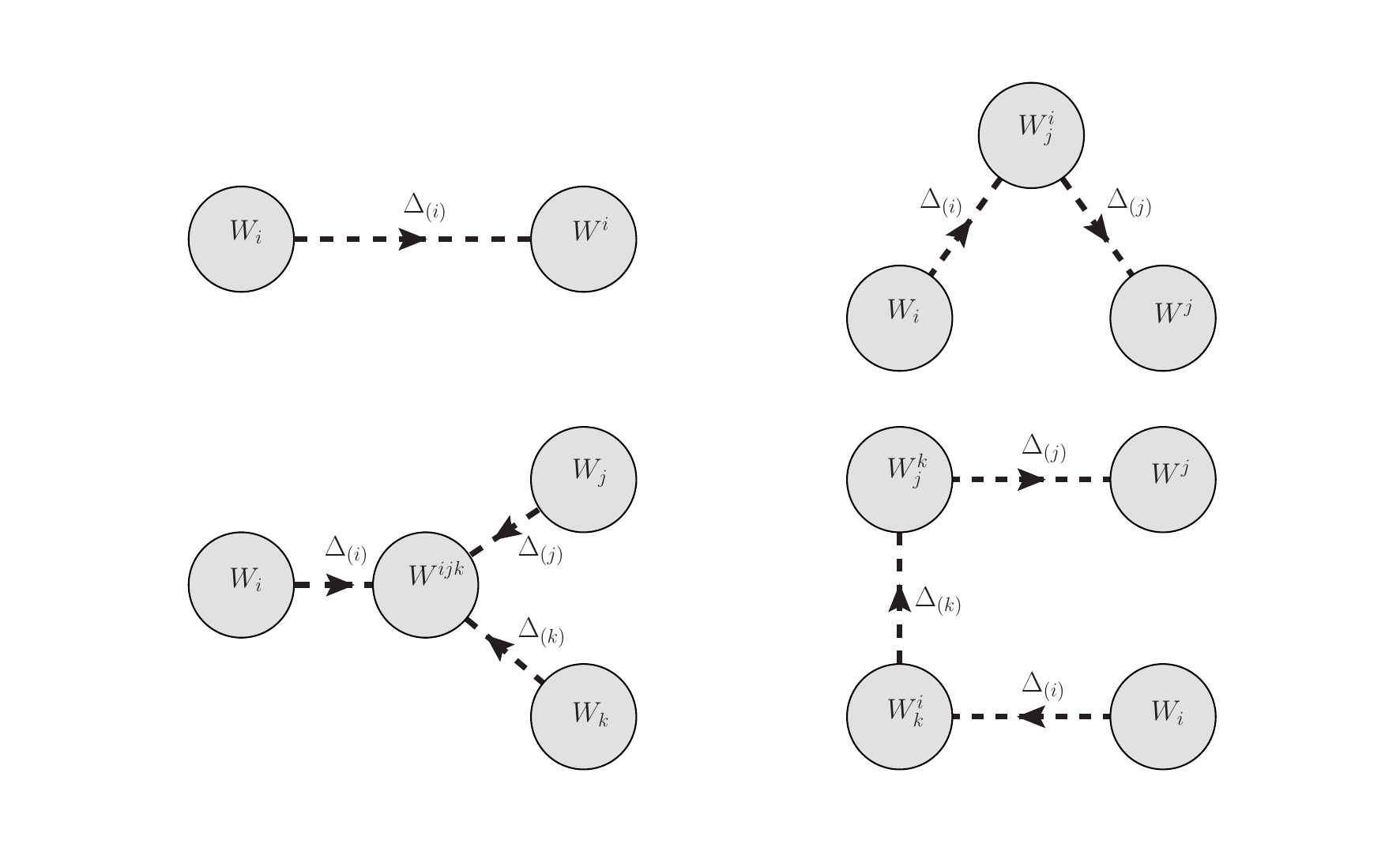}
\vspace{-1cm}
\caption{
Feynman diagrams contributing to the effective Lagrangian to order $1/M^2$. Non-equivalent permutations of the arrow directions shown here should be considered as well.}
\label{fig:diagrams}
\end{figure}

The integration at the tree level can be performed solving the classical equations of motion for the heavy fields and inserting the solutions into the original Lagrangian. This procedure manifestly preserves the gauge invariance of the original theory. Let $\sigma^i$ be each of the scalar fields. Different values of the index $i$ label different scalars, in the same or in different representations. We will use upper and lower indices for the fields and their complex conjugates, respectively. The covariant propagator is
\be
\Delta_i=- \left(D_i^2+M^2_i \right)^{-1} = -\frac{1}{M_i^2}\left(1-\frac{D_i^2}{M_i^2}\right) + O(1/M^6) , \label{propagator}
\ee
with $D_i$ the covariant derivative acting on $\sigma^i$. The part of the Lagrangian that contains the extra scalars reads
\be
\eta_{(i)} \sigma^\dagger_i \Delta_{(i)}^{-1} \sigma^i + {\cal L}_{\mathrm{int}}. \label{scalarLag}
\ee
We are using the convention of repeated indices, contracting upper and lower indices. A parenthesis indicates an index that can run but does not count as repeated to induce the running (indices in parenthesis actually refer to the diagonal elements of a diagonal matrix, the propagator).
The interaction Lagrangian is a polynomial in $\sigma$ of degree 4 with no constant term. Hence, it will be of the form
\be
{\cal L}_{\mathrm{int}} = - \sum_{m+n=1}^4  \sigma^\dagger_{j_1}\cdots \sigma^\dagger_{j_n} W_{i_1\ldots i_m}^{j_1 \ldots j_n} \,  \sigma^{i_1}\cdots \sigma^{i_m} ,  \label{expandedLint}
\ee
where $m$ and $n$ vary independently subject to the indicated constraint.
The operators  $W_{i_1\ldots i_m}^{j_1 \ldots j_n}$ are formed with SM fields only and obey the obvious hermiticity conditions inherited from ${\cal L}_{\mathrm{int}}$. In general, they carry reducible representations of the gauge group, but the operators with one index $i$ belong to the irreducible representation of the associated $\sigma^i$ or $\sigma^\dagger_i$.
The equations of motion for the new scalars read
\be
\sigma^i = -\Delta_{(i)} \frac{\partial {\cal L}_{\mathrm{int}}}{\partial \sigma^\dagger_i}.
\label{eom} 
\ee
Note that the right-hand sides depend on the fields $\sigma$, so these are not explicit solutions. But the equations can be solved iteratively. 
The crucial point is that $\Delta_i$ is $O(1/M^2)$, whereas the couplings in ${\cal L}_{\mathrm{int}}$ are either dimensionless or $O(M)$, at most.
Therefore, the iterative solution of (\ref{eom}) starts at $O(1/M)$, and each correction (step of the iteration) is suppressed by at least another $1/M$ factor. Using (\ref{expandedLint}) in the right-hand side of (\ref{eom}), we can perform the iteration explicitly. Then we plug the iterative solution in (\ref{scalarLag}) and~(\ref{expandedLint}), neglecting $O(1/M^3)$ terms, to obtain 
\begin{align}
{\cal L}_{\mathrm{eff}} = & \; {\cal L}_{\mathrm{SM}}  - \Delta_{(i)} W_i W^i - \Delta_{(i)} \Delta_{(j)} \left(W_{ij} W^i W^j + W^{ij} W_i W_j + W^i_j W_i W^j \right) \nn
& \mbox{} - \Delta_{(i)} \Delta_{(j)} \Delta_{(k)} \left( W^{ijk} W_i W_j W_k + W_{ijk} W^i W^j W^k + W^{ij}_k W_i W_j W^k + W^i_{jk} W_i W^j W^k \right. \nn
&  \left. \mbox{}+ 2 W^{ik} W^j_k W_i W_j + 2 W_{ik} W^k_j W^i W^j + 4 W^{ik}W_{kj} W_iW^j + W^i_k W^k_j W_i W^j  \right) \nn
  & \mbox{} + O(1/M^3).  \label{integratedLint}
\end{align}
In each term, $\Delta_{(i)}$ acts on the operator with an upper index $i$. The covariant propagators $\Delta_i$ are to be expanded in $1/M_i^2$, as in (\ref{propagator}).
This result matches precisely the one obtained from the possible Feynman diagrams with heavy-scalar propagators that contribute to order $1/M^2$, shown in Fig.~\ref{fig:diagrams}. The blobs in this figure represent the SM operators $W^{i_1\ldots i_m}_{j_1\dots j_n}$ with $m$ incoming and $n$ outgoing lines, and the arrowed lines represent the covariant propagators $\Delta_i$.

Observe that in all the terms in (\ref{integratedLint}) except the
ones in the last line, there is one $W_i$ or $W^i$ operator for each
propagator index $i$. $W_i$  and $W^i$ arise from terms in ${\cal
  L}_{\mathrm{int}}$ with only one heavy scalar ($\sigma^i$ or
$\sigma^\dagger_i$), which are what we have called linear
interactions. In the last line (corresponding to the last Feynman
diagram), on the other hand, all terms have a propagator with index
$k$ that is not attached to any $W_k$ or $W^k$, but only to operators
with two indices. However, these terms are actually the contraction of
the two one-index operators $\widetilde{W}_k=[W_{kj} \, \Delta_{(j)}
  W^j]$ and $\widetilde{W}^k=[W^{kj} \, \Delta_{(j)} W^j]$ (or
variations in the position of the indices). The operators
$\widetilde{W}_k$ and $\widetilde{W}^k$ are in the same gauge
representation as $W_k$ and $W^k$, respectively. Moreover, to allow
for an $O(1/M^2)$ contribution, the operators $\widetilde{W}_k$ and
$\widetilde{W}^k$ must have a dimensionless coefficient and, hence,
scaling dimension four. Therefore, the scalars $\sigma^k$
($\sigma^\dagger_k$) associated to $\widetilde{W}_k$
($\widetilde{W}^k$) must also belong to a representation that allows
for renormalizable linear interactions. We conclude that only the scalar fields in the irreducible representations of Table~\ref{table:newscalars} contribute at the tree level to the effective Lagrangian to dimension six. 

Note also that the last two topologies in Fig.~\ref{fig:diagrams} only contribute to this order when the four blobs contain $O(M)$ dimensionful couplings, which requires that all of them arise from trilinear terms in the scalar potential. In particular, the SM fermions only appear through the diagrams with the first and second topologies. Gauge bosons only arise at dimension six  from the covariant propagator of the first topology, when both blobs represent trilinear interactions in the potential.

A complete basis of gauge-invariant operators to dimension six, including the ones generated by heavy scalars, is given in Appendix~A. The relevant interactions of arbitrary scalars and the detailed results of the integration are collected in Appendix~\ref{app: NS_OpCoeff}.\footnote{In some cases, we employ algebraic identities and/or field redefinitions to transform the induced operators into the ones in our basis.} 

%------------------------------------------------------------------------------------------

\section{Observable effects of new scalars}
\label{section_ObsEff}

The leading indirect effects of new heavy scalar particles on physical
observables are described by the dimension five and six effective
operators, with coefficients given
in~tables~\ref{Table:S0Table}-\ref{Table:MultiScalTable} in
Appendix~\ref{app: NS_OpCoeff}. In this section we give an overview of
these effects. We discuss colored and colorless scalars in turn. 

The scalar fields with $SU(3)_c$ quantum numbers manifest themselves,
at dimension six, only through four-fermion interactions. Therefore they
can be tested in two-to-two fermion processes or in particle decays. The
flavor structure of the scalar interactions makes them particularly
sensitive to constraints from flavor-violating processes, although it
is always possible to go into an alignment limit in which each new scalar
couples exclusively to certain fermion generations, up to factors of
the Cabibbo-Kobayashi-Maskawa (CKM) matrix. In such a case, the bounds
from flavor-preserving processes can become dominant. We will only
consider this scenario in the following.

The scalar fields  $\omega_1$, $\omega_4$, $\Pi_1$, $\Pi_7$ and
$\zeta$ carry quantum numbers that allow for lepton-quark Yukawa
interactions. These {\it scalar leptoquarks} can be tested in the
dilepton processes $e^+ e^-\rightarrow \mt{had}$ at LEP2 and $p p
\rightarrow \ell^+\ell^-$ at the LHC, where the new particles are
exchanged in the $t$ channel, and also in low-energy experiments
(e.g. parity violation in atoms). The multiplets $\omega_1$,
$\omega_4$ and $\zeta$ admit purely hadronic interactions too. The
simultaneous presence of all these interactions introduces a violation
of lepton ($L$) and baryon ($B$) number, as is manifest by the
generation of the operators labeled as ``\BLOp'' in
Tables~\ref{Table:om4Table},~\ref{Table:om1Table}, and
\ref{Table:ZetaTable}. These contributions are proportional to the
product of one lepton-quark and one quark-quark Yukawa coupling, so
the strong constraints set by the non-observation of proton decay
forces one of these two couplings to be very small (see for
instance~\cite{Dorsner:2012nq}).  

The remaining colored representations, $\omega_2$, $\Omega_1$, $\Omega_2$, $\Omega_4$, $\Upsilon$ and $\Phi$, only admit quark Yukawa interactions and therefore generate only four-quark contact interactions. Aside from flavor observables, these can be tested in dijet production at hadron colliders. If coupled to the first and third family, top pair production (with opposite or same sign) is also possible. 

The phenomenology of colorless scalars is significantly richer. Again,
the exchange of some colorless heavy scalars generates four-fermion
operators, which in this case include operators with four leptons. In
particular, all the four-fermion interactions induced by the
multiplets ${\cal S}_{1,2}$ and $\Xi_1$ are purely leptonic. The
effects of these operators could show up in the $e^+ e^-\rightarrow
\ell^+ \ell^-$ data taken at LEP2 or in low-energy experiments such as
measurements of parity violation in M{\o}ller
scattering.~\footnote{Four-lepton operators not involving electrons
  could in principle be accessible through four-lepton production at
the LHC or ILC. The sensitivity is however only marginal if there is
no resonant production~\cite{delAguila:2014soa}.} In general, these
four-lepton operators also contribute to very sensitive lepton flavor
violating processes, such as $\mu^-\rightarrow e^+ e^- e^-$ (note that
this cannot be mediated by ${\cal S}_1$) or $\tau^-\rightarrow \mu^-
e^+ e^-$. But similarly to the quark case, these dangerous effects are
absent when the scalar couplings are properly aligned with the SM
lepton flavors.

Colorless scalars can also have visible effects in other types of
observables. 
To start with, the hypercharge-one iso-triplet $\Xi_1$ is the only scalar multiplet that can produce lepton-number violation. Indeed, it contributes to the dimension five Weinberg operator~\cite{Weinberg:1979sa} ${\cal O}_5=\overline{l_L^c} \tilde{\phi}^* \tilde{\phi}^\dagger l_L$, which generates Majorana masses for the SM neutrinos. This is nothing but the well-known seesaw mechanism of type II.  Unless the scalar iso-triplet is very heavy (and thus does not contribute to other observables), the smallness of neutrino masses requires that either the lepton Yukawa couplings of the scalar or its linear interactions with the Higgs be tiny~\cite{delAguila:2007ap}.

Colorless scalars contribute to the following three purely bosonic
dimension-six operators:
${\cal O}_{\phi \Box}=\left(\phi^\dagger \phi\right)\Box
\left(\phi^\dagger \phi\right)$, 
${\cal O}_{\phi D}=\left(\phi^\dagger D_\mu \phi\right) \left(D^\mu
\phi^\dagger \phi\right)$   
and ${\cal O}_\phi=\left(\phi^\dagger \phi\right)^3/3$.
The first one, ${\cal O}_{\phi D}$, is in one to one correspondence with the
Peskin-Takeuchi oblique $T$ parameter \cite{Peskin:1991sw}, 
\be
T=-\frac{\alpha_{\phi D}}{2\alpha_{em}}\frac{v^2}{\Lambda^2},
\ee
which is very strongly constrained by electroweak precision data (EWPD),
especially now that the value of the Higgs mass is known. Only the colorless iso-triplets
$\Xi_0$ and $\Xi_1$ contribute to $O_{\phi D}$ at the tree level. Hence, these are the 
only scalars whose effects break custodial symmetry at tree level, to dimension six in the
effective Lagrangian expansion. In this regard, note that, unlike the weak iso-singlets and
iso-doublets, quadruplets do also break custodial isospin at the tree level if they acquire a vev. 
However, because of the absence of trilinear interactions with two Higgs fields, this vev is 
suppressed by a factor of $O(1/M^2)$. Therefore, quadruplet custodial isospin breaking effects
appear starting at dimension eight.

Both ${\cal O}_{\phi D}$ and ${\cal O}_{\phi \Box}$ renormalize the
wavefunction of the physical Higgs field, 
\be
H\rightarrow \left(1 + \alpha_{\phi \Box}\frac{v^2}{\Lambda^2}-\frac 14 \alpha_{\phi D}\frac{v^2}{\Lambda^2}\right) H,
\ee
and therefore enter in most Higgs observables. 
Nevertheless, the constraints on the coefficient of ${\cal O}_{\phi
  D}$ from EWPD are significantly stronger than the ones from Higgs
physics~\cite{deBlas:2014ula,Ellis:2014jta}, and completely dominate
in the global fit. 
The operator ${\cal O}_\phi$
introduces corrections
to the Higgs vev and mass parameters, 
which can always be absorbed in the physical values. It also gives
a direct contribution to the self-coupling of the physical Higgs, 
\be
\Delta {\cal L}_{H^3}= \frac{5}{6} \alpha_\phi\frac{v^3}{\Lambda^2} H^3, 
\ee
which is in principle observable in Higgs pair production. However, the
LHC data at 8 TeV are not sensitive enough to probe the Higgs self-coupling.
Given the small cross sections for Higgs pair production in the SM, a measurement
of the Higgs self-coupling seems to be challenging even with the results of Run 2, and 
may require of the high-luminosity upgrade of the LHC \cite{Baglio:2012np}.
On the other hand, a relatively large enhancement in diHiggs production due to 
the effect of ${\cal O}_\phi$ could unveil the presence of new physics effects
before sensitivity to the SM coupling is attained.

With the exception of ${\cal S}_1$ and ${\cal S}_2$, all the colorless
multiplets contribute to Higgs observables (the singlet ${\cal S}$
and the two quadruplets $\Theta_1$ and $\Theta_3$ {\em only}
contribute to Higgs observables). 
The singlet ${\cal S}$
and the iso-triplets $\Xi_0$ and $\Xi_1$ contribute to 
${\cal O}_{\phi \Box}$ and their trilinear couplings 
can therefore be constrained by Higgs
measurements. The latter two, however, contribute to 
${\cal O}_{\phi D}$ via the same trilinear coefficients, which are
therefore constrained 
mainly by EWPD.
The quadruplets $\Theta_1$ and $\Theta_3$ only
contribute to ${\cal O}_\phi$, so, to dimension six, their effects are not observable in
current data. 

Finally, the scalar-fermion operators  ${\cal O}_{e\phi}=\left(\phi^\dagger \phi\right)\left(\overline{l_L}\phi e_R\right)$, ${\cal O}_{d\phi}=\left(\phi^\dagger \phi\right)\left(\overline{q_L}\phi d_R\right)$, and ${\cal O}_{u\phi}=\left(\phi^\dagger \phi\right)\left(\overline{q_L}\tilde{\phi} u_R\right)$ correct the SM Yukawa interactions:
\be
\Delta {\cal L}_{\mt{Yukawa}}=\frac{1}{\sqrt{2}}H\left( \left(\alpha_{e\phi}\right)_{ij} \overline{e_L^i} e_R^j +\left(V \alpha_{u\phi}\right)_{ij} \overline{u_L^i} u_R^j +\left(\alpha_{d\phi}\right)_{ij} \overline{d_L^i} d_R^j+ \hc\right)\frac{v^2}{\Lambda^2}.
\ee
In this equation we have already reabsorbed the corrections to the masses in the definition of the SM Yukawa matrices. These operators are generated by the colorless iso-triplets and iso-doublet. For $\Xi_0$ and $\Xi_1$, the coefficients are proportional to the (squared) trilinear couplings, which as pointed out above contribute to the $T$ parameter and are strongly constrained by EWPD. In the case of the doublet $\varphi$, the coefficients are proportional to the product of the scalar coupling $\lambda_\varphi \left(\varphi^\dagger \phi\right)\left(\phi^\dagger \phi\right) + \hc$ and the corresponding fermionic coupling. The Higgs observables only constrain this product but not the individual couplings. (The scalar coupling enters quadratically in the coefficient of ${\cal O}_\phi$ but, as indicated above, there is no significant bound on this coefficient at present.) 

In the next section we give some numerical results for bounds that can be obtained on the couplings and masses of the heavy scalar particles from the available measurements.

%------------------------------------------------------------------------------------------

\section{Precision constraints on new scalars}
\label{section_Pheno}

As explained in the previous section, the effects of the scalar couplings in Tables \ref{Table:S0Table}-\ref{Table:PhiTable} can potentially be observed in several different physical processes. 
The good agreement of the SM predictions with most of the current observations implies bounds on the different interactions.
In this section, we use the effective Lagrangian results obtained in
the previous sections to derive flavor-conserving limits on some of
the scalar representations.  

For the sake of simplicity, in the fits presented here we consider
only one scalar multiplet at a time and always assume that only one of
the possible couplings of each scalar is
non-vanishing.\footnote{Besides the obvious simplifications of
  reducing the number of free parameters and allowing for a simple
  one-dimensional presentation of the results, this assumption allows
  us to use the $pp\rightarrow j j$ results of
  \cite{Domenech:2012ai}. Indeed, the four-quark operators ${\cal
    O}_{qud}^{(1)}$ and ${\cal O}_{qud}^{(8)}$, not considered in that
    reference, are generated for some scalar representations, but their coefficients
    always involve the product of two different Yukawa couplings.} 
In most cases, this assumption gives rise to conservative limits. More general scenarios are certainly interesting and can be studied with the tools provided in this paper. At any rate, it is important to observe that there are strong phenomenological reasons for not considering certain couplings simultaneously, as we explain next.

As stressed above, the new scalar fermionic interactions do not
conserve flavor in general. Thus, they are subject to the constraints
imposed by observables measured in flavor-violating processes, which
are usually much stronger than the ones from flavor-conserving
observables. For pure hadronic interactions in the form of four-quark
operators, for instance, the observables with $\Delta F=2$ transitions
(e.g. $\epsilon_K$ or $\Delta m_K$, measured in $K^0-\overline{K^0}$
mixing) impose bounds on the new physics scale typically around
$10^2$-$10^4$ TeV, assuming order-one
couplings~\cite{Isidori:2010kg}. Lepton flavor violating processes
also impose strong bounds, especially from rare decays such as
$\mu^-\rightarrow e^+ e^- e^-$ or $\mu\rightarrow
e\gamma$. Flavor-preserving results are meaningful in scenarios in which flavor constraints are subdominant or do not apply. For instance, flavor constraints can be avoided in a natural manner by enforcing an appropriate symmetry on these SM extensions, which requires extending each scalar gauge multiplet to a full multiplet under the corresponding flavor group~\cite{Grinstein:2011dz}. From the point of view of our model-independent description of new scalars, each of these flavor multiplets corresponds to several copies of one of our gauge-covariant multiplets, with correlated couplings.\footnote{Since Ref.~\cite{Grinstein:2011dz} concentrates on quark processes, its classification does not include the scalar representations that do not have purely quark interactions.} The presence of tree-level flavor changing neutral currents can also be softened if the new Yukawa interactions are adequately aligned with the SM flavors.
In particular, any flavor violation can be removed---up to terms
suppressed by the corresponding CKM matrix elements---if, for each
scalar multiplet, only one entry of the new Yukawa matrices is
non-zero. We will restrict ourselves to this case in the present
section. This tuned 
choice provides conservative bounds. It also helps to establish in
which places certain new physics effects might be hidden and to
determine their maximum size allowed by current data. 

We also noted in the discussion of the previous section that in several cases the new scalars can contribute to other extremely sensitive physical observables, such as proton decay or neutrino masses. Since one can always assign definite $B$ and $L$ numbers to the new scalars, such contributions can only appear as the product of two interactions selecting different assignments of these quantum numbers. 
Therefore, they are always avoided when only one of these couplings is non-vanishing.  

Finally, certain contributions to four fermion operators that would give rise to charged-current interactions mediating rare decays are also absent when we only consider one non-zero coupling. For instance, the observable $R_\pi=\Gamma(\pi^+\rightarrow\nu e^+)/\Gamma(\pi^+\rightarrow\nu\mu^+)$ set bounds on the operators ${\cal O}_{qde}$, ${\cal O}_{ledq}$ and ${\cal O}_{luqe}$, which are significantly stronger than the ones from the LHC and EWPD considered here \cite{Carpentier:2010ue,deBlas:2013qqa}. The same holds for same-sign top pair production~\cite{AguilarSaavedra:2011zy}.

In the scenario we are considering, with no contribution to any of
these sensitive observables, the most relevant constraints on the
couplings and masses of the new scalars come from flavor-, $B$- and
$L$-blind observables. Our fits combine the bounds on dimension six
interactions from EWPD~\cite{Blas:2013ana}\footnote{This includes the usual $Z$-pole data~\cite{ALEPH:2005ab}, $\Delta \alpha_\mt{had}^{(5)}(M_Z^2)$~\cite{Davier:2010nc}, $\alpha_s(M_Z^2)$~\cite{Agashe:2014kda}, the top~\cite{ATLAS:2014wva} and Higgs~\cite{Aad:2014aba} masses, the $W$ mass and width~\cite{Group:2012gb}, the final LEP2 results of $e^+ e^-\rightarrow \bar{f}f$~\cite{Schael:2013ita}, unitarity constraints on the the CKM matrix~\cite{Agashe:2014kda}, as well as several low-energy measurements~\cite{Agashe:2014kda}.}, LHC dilepton~\cite{deBlas:2013qqa} and dijet searches~\cite{Domenech:2012ai}, and measurements of Higgs observables~\cite{deBlas:2014ula}.\footnote{We do not include here limits from (opposite sign) top pair production on couplings mixing the first and third generation of quarks. These can be obtained from LHC data and the results in~\cite{AguilarSaavedra:2010zi} and~\cite{AguilarSaavedra:2011vw} and will be considered elsewhere.}
In all the analyses we fix the SM inputs to their best-fit values in the absence of extra scalars,
\be
m_H=125.1\pm0.2\units{GeV},~~m_t=173.8\pm0.8\units{GeV},~~M_Z=91.1880\pm0.0020\units{GeV},\nonumber\ee
\be
\alpha_s(M_Z^2)=0.1186\pm0.0006,~~\Delta \alpha_{\mt{had}}^{(5)}(M_Z^2)=0.02754\pm0.00010,
\ee
and vary only the new-physics parameters. This is a good approximation, since large effects are not allowed.
The limits we obtain in this way are presented in
Table~\ref{table_OneSLimitsNoColor} for the colorless multiplets, and
in Table~\ref{table_OneSLimitsColor} for the ones charged under
$SU(3)_c$. In all cases the limits apply to ratios of couplings and
masses, which are the quantities that appear in the coefficients of
the effective operators. (In some cases tailored searches can give
better bounds when the new scalars can be directly
produced~\cite{Dorsner:2014axa}.)  

% Table 95% limits (Colorless scalars)
\begin{table}[t]
\begin{center}
{\footnotesize
\begin{tabular}{c c c c}
\ctoprule
Scalar&~~Parameter~~&~~95\% C.L. Bound~~\\
          & &[TeV$^{-1}$]\\
\cmrule
&&\\[-0.4cm]
${\cal S}$&$\frac{\left|\kappa_{{\cal S}}\right|}{M_{{\cal S}}^2}$&$1.55$\\
&&\\[-0.4cm]
\crowcolor&&\\[-0.4cm]
\crowcolor${\cal S}_1$&$\frac{\left|y_{{\cal S}_1}^{l}\right|}{M_{{\cal S}_1}}$&$\left(\begin{array}{c c c} -&0.08&-\\ 0.08&-&-\\ -&-&-\end{array}\right)$\\
&&\\[-0.4cm]
${\cal S}_2$&$\frac{\left|y_{{\cal S}_2}^{e}\right|}{M_{{\cal S}_2}}$&$\left(\begin{array}{c c c} 0.36&0.19&0.28\\ 0.19&-&-\\ 0.28&-&-\end{array}\right)$\\
&&\\[-0.4cm]
\crowcolor&&\\[-0.4cm]
\crowcolor$\varphi$&$\frac{\left|y_{\varphi}^{e}\right|}{M_{\varphi}}$&$\left(\begin{array}{c c c} 0.26&0.56&0.79\\ 0.56&-&-\\ 0.79&-&-\end{array}\right)$\\
\crowcolor&$\frac{\left|\left(y_{\varphi}^{d}\right)_{11}\right|}{M_{\varphi}}$&$0.61$\\
\crowcolor&$\frac{\left|\left(y_{\varphi}^{u}\right)_{11}\right|}{M_{\varphi}}$&$0.44$\\
&&\\[-0.4cm]
$\Xi_0$&$\frac{\left|\kappa_{\Xi_0}\right|}{M_{\Xi_0}^2}$&$0.11$\\
&&\\[-0.4cm]
\crowcolor&&\\[-0.4cm]
\crowcolor$\Xi_1$&$\frac{\left|\kappa_{\Xi_1}\right|}{M_{\Xi_1}^2}$&$0.04$\\
\crowcolor &$\frac{\left|y_{\Xi_1}^{l}\right|}{M_{\Xi_1}}$&$\left(\begin{array}{c c c} 0.33&0.09&0.18\\ 0.09&-&-\\ 0.18&-&-\end{array}\right)$\\
&&\\[-0.4cm]
\cbottomrule
\end{tabular}
\caption{
Bounds on the colorless new scalars from flavor-preserving
  observables.  The
  results for the Yukawa matrices are obtained from a fit to each one of
  the entries of the coupling matrices at a time. The limit on
  $\kappa_{\cal S}$ is determined exclusively by the Higgs data, while
  the ones on the $\kappa_{\Xi_i}$ couplings are dominated by the EWPD
  limits on the $T$ parameter. Leptonic couplings are constrained by
  the LEP2 ($e^+ e^- \rightarrow \ell^+\ell^-$) and low energy
  measurements (e.g. M{\o}ller and $\nu$-electron scattering), while
  the hadronic ones are bounded by the LHC dijet angular
  distributions. 
\label{table_OneSLimitsNoColor}}
}
\end{center}
\end{table}

Let us comment on the few absences in those tables. In the colorless case, we cannot put meaningful bounds on the quadruplet couplings. As explained in the previous section, they only modify the Higgs self-coupling, which is not significantly constrained by the LHC data at 8 TeV. In the colored sector, we have not presented any bounds for $\omega_2$, nor for the hadronic couplings of $\omega_4$ and $\zeta$. These could be in principle constrained by the LHC dijet data. However, the hadronic couplings of these three multiplets are antisymmetric and necessarily involve more than one family. Hence, they go beyond the first-family approximation used in~\cite{Domenech:2012ai}. Putting bounds on them would require an extended analysis.

% Table 95% limits (Colored scalars)
\begin{table}[t]
\begin{center}
{\footnotesize
\begin{tabular}{c c c}
\ctoprule
\!Scalar\!&~Parameter~&~95\% C.L. Bound~\\
          & &[TeV$^{-1}$]\\
\cmrule
\crowcolor&&\\[-0.2cm]
\crowcolor$\omega_1$&$\frac{\left|y_{\omega_1}^{ql}\right|}{M_{\omega_1}}$&$\left(\begin{array}{c c c} 0.19&0.53&-\\ 0.40&-&-\\ -&-&-\end{array}\right)$\\
\crowcolor&$\frac{\left|\left(y_{\omega_1}^{qq}\right)_{11}\right|}{M_{\omega_1}}$&$0.24$\\
\crowcolor&$\frac{\left|y_{\omega_1}^{eu}\right|}{M_{\omega_1}}$&$\left(\begin{array}{c c c} 0.27&0.49&-\\ 0.48&-&-\\ -&-&-\end{array}\right)$\\
\crowcolor&$\frac{\left|\left(y_{\omega_1}^{du}\right)_{11}\right|}{M_{\omega_1}}$&$0.47$\\
&&\\[-0.2cm]
$\omega_4$&$\frac{\left|y_{\omega_4}^{ed}\right|}{M_{\omega_4}}$&$\left(\begin{array}{c c c} 0.28&0.98&0.98\\ 0.42&-&-\\ -&-&-\end{array}\right)$\\
&&\\[-0.2cm]
\crowcolor$\Pi_1$&$\frac{\left|y_{\Pi_1}^{ld}\right|}{M_{\Pi_1}}$&$\left(\begin{array}{c c c} 0.27&1.80&1.80\\ 0.48&-&-\\ -&-&-\end{array}\right)$\\
\crowcolor&&\\
\crowcolor&&\\
\cbottomrule
\end{tabular}
\quad
\quad
\begin{tabular}{c c c}
\ctoprule
\!Scalar\!&~Parameter~&~95\% C.L. Bounds~\\
          & &[TeV$^{-1}$]\\
\cmrule
\crowcolor&&\\[-0.4cm]
\crowcolor$\Pi_7$&$\frac{\left|y_{\Pi_7}^{lu}\right|}{M_{\Pi_7}}$&$\left(\begin{array}{c c c} 0.27&1.04&-\\ 0.33&-&-\\ -&-&-\end{array}\right)$\\
\crowcolor &$\frac{\left|y_{\Pi_7}^{eq}\right|}{M_{\Pi_7}}$&$\left(\begin{array}{c c c} 0.29&0.93&1.06\\ 0.32&-&-\\ -&-&-\end{array}\right)$\\
&&\\[-0.4cm]
$\Omega_1$&$\frac{\left|\left(y_{\Omega_1}^{ud}\right)_{11}\right|}{M_{\Omega_1}}$&$0.78$\\
\crowcolor&&\\[-0.4cm]
\crowcolor$\Omega_2$&$\frac{\left|\left(y_{\Omega_2}^{d}\right)_{11}\right|}{M_{\Omega_2}}$&$0.68$\\
&&\\[-0.4cm]
$\Omega_4$&$\frac{\left|\left(y_{\Omega_4}^{u}\right)_{11}\right|}{M_{\Omega_4}}$&$0.47$\\
&&\\[-0.4cm]
\crowcolor&&\\[-0.4cm]
\crowcolor$\zeta$&$\frac{\left|y_{\zeta}^{ql}\right|}{M_{\zeta}}$&$\left(\begin{array}{c c c} 0.21&0.30&-\\ 0.66&-&-\\ 0.47&-&-\end{array}\right)$\\
&&\\[-0.4cm]
$\Phi$&$\frac{\left|\left(y_{\Phi}^{qu}\right)_{11}\right|}{M_{\Phi}}$&$0.88$\\
&$\frac{\left|\left(y_{\Phi}^{dq}\right)_{11}\right|}{M_{\Phi}}$&$1.12$\\
&&\\[-0.4cm]
\crowcolor&&\\[-0.4cm]
\crowcolor$\Upsilon$&$\frac{\left|y_{\Upsilon}^{q}\right|}{M_{\Upsilon}}$&$0.32$\\
&&\\[-0.45cm]
\cbottomrule
\end{tabular}
\caption{
Bounds on the colored new scalars from flavor-preserving
  observables. The results for the Yukawa matrices are obtained from a
  fit to each one of the entries of the coupling matrices at a time. All these interactions are constrained by two to two fermion processes. Leptoquark interactions are bounded by LEP 2 $e^+ e^- \rightarrow \mt{had}$ data, low energy measurements (e.g. Atomic parity violation, $\nu$-nucleon scattering), CKM unitarity, and dilepton searches at the LHC. Purely hadronic bounds are again obtained only from the LHC $pp\rightarrow jj$ angular distributions. See text for more details. 
\label{table_OneSLimitsColor}}
}
\end{center}
\end{table}

Finally, let us discuss the range of validity of the effective Lagrangian. In this approach, the results are given on the ratios $y_{\sigma}/M_\sigma$, where $y_\sigma<4\pi$ to allow for a loop expansion, or $\kappa_{\sigma}/M_\sigma^2$, where we assumed $\left|\kappa_\sigma\right| \lesssim M_\sigma$. To guarantee the validity of our bounds, we need to assume that the
scalar masses are sufficiently larger than the relevant energies and
momenta of the processes we consider. This condition depends on each
observable and coupling and is always satisfied by large enough values
of the masses. But for large masses the upper region of the allowed
parameters may involve strong couplings that threaten perturbativity
and thus the validity of the tree-level approximation. This can happen
when the limits on coupling/mass ratios are weak. One example is the
scalar singlet ${\cal S}$, which is only observable through its
contributions to the Higgs boson wave function via the operator ${\cal
  O}_{\phi \Box}$, with mild
limits~\cite{deBlas:2014ula,Ellis:2014jta}. For $\left|\kappa_{\cal
  S}\right|\sim M_{{\cal S}}$, the bound in
Table~\ref{table_OneSLimitsNoColor} implies $M_{{\cal S}}\gtrsim 700
\units{GeV}$, which is close to the scale probed at the LHC, and the
validity of this bound might be questioned. In some cases, entries
involving electrons coupling with the second and third family of
quarks are also relatively weak $\sim {\cal
  O}(1$-$2)\units{TeV}^{-1}$. For weakly coupled scenarios
($y_\sigma<1$) this implies the new scalar masses can be around
500-1000 GeV. However, these entries are only constrained by the LEP 2
data, which involves lower energies $\sqrt{s}\leq209\units{GeV}$. Those entries that can modify
dilepton production at the LHC can be as large as
$\sim0.5\units{TeV}^{-1}$ for the case of $\Pi_1$. This translates
into a mass scale of $\sim 2\units{TeV}$, which is in principle
accessible by the LHC at $\sqrt{s}=8$ TeV. However, dilepton processes
at the LHC can only be mediated by leptoquarks in $t$-channel. Thus,
even for such values of masses and weak couplings, the effective theory
remains valid in a large region of the phase space and can give
a good approximation to the integrated observables (see the quantitative discussion of an analogous process in~\cite{AguilarSaavedra:2011vw}). Finally, the
results for the pure hadronic interactions should be taken with
care. Again, for order one couplings the corresponding mass scales can
be relatively small, while the dijet angular distributions used in
\cite{Domenech:2012ai} to set bounds correspond to dijet masses
$M_{jj}> 3\units{TeV}$. Unlike the dilepton case, these scalars can be
produced in $s$-channel. For instance, for the case with the weakest
bound, the color-octet iso-doublet $\Phi$, demanding $M_\Phi
>3.9\units{TeV}$ ( the highest $M_{jj}$ value observed in the CMS
analysis \cite{Chatrchyan:2012bf} used in \cite{Domenech:2012ai}), the
hadronic couplings needed to saturate the bounds must be
$(y_\Phi^{dq})_{11}\gtrsim 4$, and one may start worrying about the
precision of the perturbative (asymptotic) series. Note, nevertheless, that for $s$-channel processes the
limits obtained with the effective Lagrangian give a conservative
estimate of the actual limit. In summary, with the current constraints
the effective Lagrangian approach provides in general a good approximation
for heavy scalars and a large range of values of their couplings. 

%------------------------------------------------------------------------------------------
\vspace{0.1cm}
\section{Scalar extensions with other new particles}
\label{section_NSNVNF}
\vspace{0.1cm}

In the previous section we have considered in detail SM extensions with only one scalar multiplet.  In this case, there are strong correlations between different observables. In this section we discuss the interplay between the effects of different scalars, and also between particles with different spin. The effective Lagrangian formalism used in this article allows an easy comparison of the effects of different sources of new physics. In particular, it helps to identify at the Lagrangian level those places where a (partial) cancellation between the virtual effects of different new particles in physical observables can take place~\cite{delAguila:2011zs, delAguila:2011yd}. This is useful for model builders to construct scenarios with not too heavy (or not too weakly coupled) particles that are consistent with the existing phenomenological constraints. Such cancellations require a large correlation between the effects of different new particles. Although they correspond to small regions in the parameter space of generic models, in some cases they can be made natural by imposing extra symmetries.

Let us first point out that, at the tree level, the dimension-six effective Lagrangian in extensions of the SM with arbitrary new particles of spin 0, 1/2 and 1 is simply the sum of the effective Lagrangian obtained here and the ones in Refs.~\cite{delAguila:2000rc,delAguila:2008pw,delAguila:2010mx}.\footnote{A direct comparison of the effective Lagrangian results presented here with those in Refs.~\cite{delAguila:2000rc,delAguila:2008pw,delAguila:2010mx} requires to perform certain field redefinitions and Fierz reorderings, since the basis employed in those works has some redundant interactions, and use different definitions for some operators. All the transformations needed to relate both bases are provided in Appendix~\ref{app: L6_basis}, where we also introduce the full basis of dimension six operators we use, which can be compared to the one in Table~7 in Ref.~\cite{delAguila:2010mx}.} Indeed, a simple extension of the argument in Section~\ref{section_Eff_Lag} shows that mixed contributions from particles of different spin only appear at higher dimension. Therefore, the effective Lagrangians in those references and in this paper completely characterize the largest effects of arbitrary extensions of the SM with new heavy particles. Note in this regard that particles of spin higher than 1 only interact via non-renormalizable couplings, which are naturally suppressed.

In what follows we study, for each type of interaction induced by the extra scalars to dimension six, the different sources of new physics (new scalars, fermions or vectors) that can cancel at the tree-level the effects from the virtual exchange of scalar bosons. We discuss the cancellations at the operator level, which is a sufficient (and often necessary) condition to guarantee the cancellation in physical observables.

\begin{itemize}

{\item {\bf Dimension-five operators:} The Weinberg operator only arises when we integrate the hypercharge-one iso-triplet $\Xi_{1}$. Having no definite sign, contributions from different triplets could cancel each other, or the ones coming from the other two possible seesaw messengers, i.e. new lepton singlets and triplets with hypercharge zero.}

{\item {\bf Oblique operators:} Cancellation between custodial isospin
  breaking contributions can occur between the two triplets, as both
  have definite, opposite sign. This is also possible with new vector
  singlets with hypercharge zero (hypercharge one), which yield negative (positive) definite contributions to $\alpha_{\phi D}$. New vector triplets with hypercharge one also yield negative contributions to this operator~\cite{delAguila:2010mx}.}

{\item {\bf Scalar operators}: These include ${\cal O}_\phi$ and ${\cal O}_{\phi \Box}$. A look at the second row of Table~\ref{Table:MultiScalTable} in Appendix~\ref{app: NS_OpCoeff} shows all the possible contributions to the operator ${\cal O}_\phi$, including collective contributions that appear when two different species of new scalars are present at the same time. Contributions to ${\cal O}_{\phi \Box}$, on the other hand, are much simpler, and are negative (positive) definite for ${\cal S}$ ($\Xi_{0,1}$). This allows for cancellations between colorless iso-singlets and iso-triplets.}

{\item {\bf Scalar-Fermion operators:} In extensions with new scalars only, these operators only appear in the case of colorless iso-doublets, $\varphi$ or iso-triplets, $\Xi_{0,1}$. However, for the latter they always arise through a field redefinition, necessary to bring all the contributions in the dimension-six Lagrangian into the chosen basis. As a consequence, the flavor structure of the coefficients of the operators ${\cal O}_{e\phi}$ and ${\cal O}_{d\phi}$, and ${\cal O}_{u\phi}$ coming from triplets is always SM-like (proportional to the SM Yukawa couplings), while the one from doublets $\varphi$ can be completely generic. Moreover, while all the contributions of arbitrary triplets have the same sign, the sign of the genuine contributions from $\varphi$ is indefinite. Therefore, cancellations between scalars are always possible.

Contributions from new vectors \cite{delAguila:2010mx}, also appear when the SM equations of motion are used. They only come from colorless hypercharge zero or one iso-singlets, or iso-triplets (${\cal B}$, ${\cal B}_1$, ${\cal W}$ and ${\cal W}_1$ in the notation of \cite{delAguila:2010mx}), and are also SM-like. 

Finally, heavy fermions can generate these contributions either after applying the SM equation of motion, if only one fermion species is present, or as a result of the combined effect of extra fermionic iso-doublets and new fermion iso-singlets or iso-triplets~\cite{delAguila:2000rc,delAguila:2008pw}. Even in the case of only one fermion, the contributions to ${\cal O}_{f\phi}$ always involve the flavor structure of the new fermionic interactions, and therefore are general a priori. Hence, some interplay with the contributions from scalar doublets is possible, although an eventual cancellation of all the scalar effects may require several different new fermion multiplets.
}

{\item {\bf Four-fermion operators:}  Upon inspection of the new
  scalar contributions to four-fermion operators, it can be seen that,
  for a fixed set of flavor indices with $i=j$,~$k=l$, the operator
  coefficients of all four-fermion interactions involving at most two
  different types of SM fermion multiplets have a definite
  sign. Although the contributions to four-fermion interactions with
  three or four different multiplets have no definite sign, they are
  always correlated with operators involving only one or two kinds of
  multiplets. Moreover, for a particular operator, contributions from
  scalars of different types have either the same sign, or are
  proportional to the contribution to another operator with the same
  field content where both scalars contribute additively. For
  instance, $\varphi$ and $\Phi$ contributions to $\alpha_{qu}^{(8)}$
  have opposite sign and can balance each other. However, each
  individual contribution is proportional to the corresponding one to
  $\alpha_{qu}^{(1)}$, where both have the same sign. 
Therefore, a complete cancellation of the effects from such couplings to two-to-two fermion processes is not possible in extensions with extra scalars only. This is quite similar to the situation for extra vectors \cite{delAguila:2010mx}. However, as illustrated in Ref.~\cite{delAguila:2011yd} for the case of pure leptonic interactions, a cancellation between the four fermion effects coming from new scalars and the ones from new vector particles is possible in many cases, although it comes at the price of a significant fine tuning. New vectors of hypercharge $Y\not=0$ contribute with a definite sign to four-fermion operators involving at most two types of SM multiplets if $i=j$,~$k=l$, exactly as in the scalar case. For vectors of zero hypercharge, only the operators where all fermions belong to the same SM representation can have a definite sign. This is always the case for $i=j=k=l$ and, in certain cases, for $i=l$,~$j=k$.\footnote{Note that in this case, the corresponding scalar contributions always have a definite sign.}
Instead of going over each operator/scalar/vector and listing all the
possible cancellations, we show in Table~\ref{SV_4Fops2} those
interactions that are common for each scalar-vector pair, indicating
the relative sign between the different contributions for both cases, $i=l$,~$j=k$~($i\not= j$), and $i=j=k=l$
(where some restrictions appear in the case of hypercharge-zero vector
fields).
In general, for a given operator with four multiplets of the same kind, one can always choose a scalar and a hypercharge-zero vector field such that, tuning the corresponding scalar/vector couplings, contributions with opposite sign are obtained.
 Table~\ref{SV_4Fops} contains the same information for the
case of four-fermion operators built from at most two types of multiplets, with $i=j$,~$k=l$~($i\not=k$).

Note that a relative minus sign between the contributions from two particles to a given operator does not always guarantee that a complete cancellation of the new physics effects is possible. The reason is that the contributions to some operators with the same field content are in many cases correlated, and a cancellation in all those operator coefficients does not usually take place at the same time. However, for each configuration of four fermionic fields, there are at most two independent operators in the dimension six basis. And, as can be seen from the tables, for each scalar (vector) field and pair of such operators, one can always find a pair of vectors (scalars) that contribute to both operators with the adequate signs to cancel the total contribution. Therefore, we conclude that for any given four-fermion process receiving contributions from one new particle through an arbitrary set of four-fermion operators, it is always possible to find a combination of new fields that, after the adequate tuning in their couplings, cancels out all the new effects. Correlations with other types of operators can be easily avoided.
}
\end{itemize}

% Cancellations: All indices equal
\begin{table}[t]
\begin{center}
{\scriptsize
\begin{tabular}{c | c c c c c c c c c c c}
\ctoprule
&${\cal S}_1$&${\cal S}_2$&$\Xi_1$&${\omega}_{1}$&${\omega}_{2}$&${\omega}_{4}$&$\zeta$&$\Omega_{1}$&$\Omega_{2}$&$\Omega_{4}$&$\Upsilon$\\
\hline
\crowcolor\!${\cal B}_\mu$\!&$+\alpha_{ll}^{(1)}$&\!$\pm\alpha_{ee}$\!&\!$-\alpha_{ll}^{(1)}$\!&$-\alpha_{qq}^{(1)}$&$+\alpha_{dd}^{(1)}$&$+\alpha_{uu}^{(1)}$&$+\alpha_{qq}^{(1)}$&$+\alpha_{qq}^{(1)}$&\!$-\alpha_{dd}^{(1)}$\!&\!$-\alpha_{uu}^{(1)}$\!&\!$-\alpha_{qq}^{(1)}$\!\\
\crowcolor$ $&$$&$$&$$&$$&$ $&$$&$$&$ $&$ $&$ $&$ $\\[-0.3cm]
\crowcolor$ $&$$(---)&$(-\alpha_{ee})$&$(-\alpha_{ll}^{(1)})$&$(-\alpha_{qq}^{(1)})$&$ $(---)&$ $(---)&$ $(---)&$ $(---)&\!$(-\alpha_{dd}^{(1)})$\!&$(-\alpha_{uu}^{(1)})$&$(-\alpha_{qq}^{(1)})$\\
$ $&$$&$$&$$&$$&$ $&$ $&$ $&$$&$ $&$ $&$$\\[-0.3cm]
\!${\cal W}_\mu$\!&$\pm\alpha_{ll}^{(1)}$&$$&\!$\pm\alpha_{ll}^{(1)}$\!&\!$\pm\alpha_{qq}^{(1)}$\!&$ $&$ $&$\pm\alpha_{qq}^{(1)}$&$\pm\alpha_{qq}^{(1)}$&$$&$$&\!$\pm\alpha_{qq}^{(1)}$\!\\
$ $&$$&$$&$$&$$&$$&$$&$ $&$ $&$$\\[-0.3cm]
$ $&$$&$$&$$&\!$\pm\alpha_{qq}^{(8)}$\!&$ $&$ $&$\pm\alpha_{qq}^{(8)}$&$\pm\alpha_{qq}^{(8)}$&$ $&$ $&\!$\pm\alpha_{qq}^{(8)}$\!\\
$ $&$$&$$&$$&$ $&$$&$$&$ $&$ $&$ $&$ $&$$\\[-0.3cm]
 &(---)&$$&\!$(-\alpha_{ll}^{(1)})$\!&\!$(+\alpha_{qq}^{(1)})$\!&$ $&$ $&$ $(---)&$ $(---)&$$&$$&\!$(+\alpha_{qq}^{(1)})$\!\\
$ $&$$&$$&$$&$$&$$&$$&$ $&$ $&$$\\[-0.3cm]
$ $&$$&$$&$$&\!$(+\alpha_{qq}^{(8)})$\!&$ $&$ $&$$(---)&$$(---)&$ $&$ $&\!$(-\alpha_{qq}^{(8)})$\!\\
$ $&$$&$$&$$&$ $&$$&$$&$ $&$ $&$ $&$ $&$$\\[-0.3cm]
\crowcolor$ $&$$&$$&$$&$ $&$ $&$ $&$ $&$ $&$$&$$&$$\\[-0.3cm]
\crowcolor\!${\cal G}_\mu$\!&$$&$$&$$&$+\alpha_{qq}^{(8)}$&$\pm\alpha_{dd}^{(1)}$&$\pm\alpha_{uu}^{(1)}$&$-\alpha_{qq}^{(8)}$&$+\alpha_{qq}^{(8)}$&$\pm\alpha_{dd}^{(1)}$&$\pm\alpha_{uu}^{(1)}$&$-\alpha_{qq}^{(8)}$\\
\crowcolor$ $&$$&$$&$$&$$&$ $&$$&$$&$ $&$ $&$ $&$ $\\[-0.3cm]
\crowcolor&$$&$$&$$&\!$(+\alpha_{qq}^{(8)})$\!&$ $(---)&$ $(---)&$ $(---)&$ $(---)&\!$(-\alpha_{dd}^{(1)})$\!&\!$(-\alpha_{uu}^{(1)})$\!&\!$(-\alpha_{qq}^{(8)})$\!\\
\!${\cal H}_\mu$\!&$$&$$&$$&\!$-\alpha_{qq}^{(1)}$\!&$ $&$ $&$+\alpha_{qq}^{(1)}$ &$+\alpha_{qq}^{(1)}$ &$$&$$&\!$-\alpha_{qq}^{(1)}$\!\\
$ $&$$&$$&$$&$$&$$&$ $&$ $&$$&$$\\[-0.3cm]
$ $&$$&$$&$$&\!$\pm\alpha_{qq}^{(8)})$&$ $&$ $&$\pm\alpha_{qq}^{(8)}$&$\pm\alpha_{qq}^{(8)}$&$$&$$&\!$\pm\alpha_{qq}^{(8)}$\!\\
 &$$&$$&$$&\!$(-\alpha_{qq}^{(1)})$\!&$ $&$ $&$ $(---)&$ $(---)&$$&$$&\!$(-\alpha_{qq}^{(1)})$\!\\
$ $&$$&$$&$$&$$&$$&$ $&$ $&$$&$$\\[-0.3cm]
$ $&$$&$$&$$&\!$(-\alpha_{qq}^{(8)})$\!&$ $&$ $&$ $(---)&$ $(---)&$$&$$&\!$(+\alpha_{qq}^{(8)})$\!\\
\cbottomrule
\end{tabular}
}
\caption{
Contributions to four-fermion interactions with $i=l$,~$j=k$~($i\not=j$), common to new scalar and (hypercharge-zero) vector fields. Only operators involving one type of multiplet, and the particles that contribute to them, are shown.  The symbols ``$+$'' (``$-$'') indicate that the contributions from scalars and vectors have the same (opposite) sign, while ``$\pm$'' indicates the absence of a definite sign in any of the sources. The same information for the case $i=j=k=l$ is provided in parenthesis. In this case, a dash (``---'') indicates the absence of contribution from the corresponding scalar particle. }
\vspace{-0.4cm}
\label{SV_4Fops2}
\end{center}
\end{table}

Summarizing, we see that the existing experimental limits are
compatible with many combinations of new particles with
sizable couplings and masses at the LHC reach. Essentially, by
including many new multiplets we are breaking the correlations in the
coefficients of the effective operators. However, in most cases the
corresponding models are too contrived and fine tuned. In the simplest
cases, symmetries may exist which make these models more natural and
appealing. The discussion in this section may be useful in the search
of such symmetries. 

% Cancellations
\begin{table}[thp]
\begin{center}
\rotatebox{90}{
\quad\quad\quad\quad{\scriptsize
\begin{tabular}{c | c c c c c c c c c c c c c c c}
\ctoprule
&\!\!${\cal S}_1$~&~${\cal S}_2$~&~$\varphi$~&~$\Xi_1$~&~${\omega}_{1}$~&~${\omega}_{2}$~&~${\omega}_{4}$~&~$\Pi_1$~&~$\Pi_7$~&~$\zeta$~&~$\Omega_{1}$~&~$\Omega_{2}$~&~$\Omega_{4}$~&~$\Upsilon$~&~$\Phi$~\\
\hline
\crowcolor${\cal B}_\mu$&$\pm\alpha_{ll}^{(1)}$&$\pm\alpha_{ee}$&$\pm\alpha_{le}$&$\pm\alpha_{ll}^{(1)}$&$\pm\alpha_{lq,ud}^{(1)}$&$\pm\alpha_{dd}^{(1)}$&$\pm\alpha_{ed}$&$\pm\alpha_{ld}$&$\pm\alpha_{lu,qe}$&$\pm\alpha_{lq}^{(1)}$&$\pm\alpha_{ud}^{(1)}$&$\pm\alpha_{dd}^{(1)}$&$\pm\alpha_{uu}^{(1)}$&$\pm\alpha_{qq}^{(1)}$&$\pm\alpha_{qu,qd}^{(1)}$\\
\crowcolor$ $&$ $&$$&$\pm\alpha_{qu,qd}^{(1)}$&$$&$\pm\alpha_{eu}$&$$&$\pm\alpha_{uu}^{(1)}$&$$&$ $&$\pm\alpha_{qq}^{(1)}$&$\pm\alpha_{qq}^{(1)}$&$$&$$&$$&$$\\
\crowcolor$ $&$ $&$$&$$&$$&$\pm\alpha_{qq}^{(1)}$&$$&$$&$$&$$&$$&$$&$$&$$&$ $&$$\\
${\cal W}_\mu$&$\pm\alpha_{ll}^{(1)}$&$$&$$&$\pm\alpha_{ll}^{(1)}$&$\pm\alpha_{lq}^{(3)}$&&$$$$&$$&$$&$\pm\alpha_{lq}^{(3)}$&$\pm\alpha_{qq}^{(1)}$&$$&$$&$\pm\alpha_{qq}^{(1)}$&$$\\
$ $&$$&$$&$$&$$&$\pm\alpha_{qq}^{(1)}$&$$&$$&$$&$$&$\pm\alpha_{qq}^{(1)}$&$-\alpha_{qq}^{(8)}$&$$&$$&$-\alpha_{qq}^{(8)}$&$$\\
$ $&$$&$$&$$&$$&$+\alpha_{qq}^{(8)}$&$$&$$&$$&$$&$+\alpha_{qq}^{(8)}$&$$&$$&$$&$$&$$\\
\crowcolor${\cal G}_\mu$&$ $&$$&$\pm\alpha_{qu,qd}^{(8)}$&$$&$\pm\alpha_{ud}^{(8)}$&$\pm\alpha_{dd}^{(1)}$&$\pm\alpha_{uu}^{(1)}$&$$&$$&$\pm\alpha_{qq}^{(8)}$&$\pm\alpha_{qq}^{(8)}$&$\pm\alpha_{dd}^{(1)}$&$\pm\alpha_{uu}^{(1)}$&$\pm\alpha_{qq}^{(8)}$&$\pm\alpha_{qu,qd}^{(8)}$\\
\crowcolor$$&$ $&$$&$$&$$&$\pm\alpha_{qq}^{(8)}$&$$&$$&$$&$$&$$&$\pm\alpha_{ud}^{(8)}$&$$&$$&$$&$$\\
${\cal H}_\mu$&$ $&$$&$$&$$&$\pm\alpha_{qq}^{(1)}$&$$&$$&$$&$$&$\pm\alpha_{qq}^{(1)}$&$\pm\alpha_{qq}^{(1)}$&$$&$$&$\pm\alpha_{qq}^{(1)}$&$$\\
$ $&$ $&$$&$$&$$&$\pm\alpha_{qq}^{(8)}$&$$&$$&$$&$$&$\pm\alpha_{qq}^{(1)}$&$\pm\alpha_{qq}^{(8)}$&$$&$$&$\pm\alpha_{qq}^{(8)}$&$$\\
\crowcolor${\cal B}^1_\mu$&$ $&$$&$$&$$&$-\alpha_{ud}^{(1)}$&$$&$$&$$&$$&$$&$-\alpha_{ud}^{(1)}$&$$&$$&$$&$$\\
\crowcolor$ $&$ $&$$&$$&$$&$+\alpha_{ud}^{(8)}$&$$&$$&$$&$$&$$&$-\alpha_{ud}^{(8)}$&$ $&$$&$$&$$\\
${\cal G}^1_\mu$&$ $&$$&$$&$$&$-\alpha_{ud}^{(1)}$&$$&$$&$$&$$&$$&$-\alpha_{ud}^{(1)}$&$$&$$&$$&$$\\
$ $&$ $&$$&$$&$$&$-\alpha_{ud}^{(8)}$&$$&$$&$$&$$&$$&$+\alpha_{ud}^{(8)}$&$$&$$&$$&$$\\
\crowcolor${\cal L}_\mu$&$ $&$$&$-\alpha_{le}$&$$&$$&$$&$$&$$&$$&$$&$$&$$&$$&$$&$$\\
${\cal U}^2_\mu$&$ $&$$&$$&$$&$-\alpha_{lq}^{(1)}$&$$&$-\alpha_{ed}$&$$&$$&$-\alpha_{lq}^{(1)}$&$$&$$&$$&$$&$$\\
$ $&$ $&$$&$$&$$&$+\alpha_{lq}^{(3)}$&$$&$$&$$&$$&$-\alpha_{lq}^{(3)}$&$$&$$&$$&$$&$$\\
\crowcolor${\cal U}^5_\mu$&$ $&$$&$$&$$&$-\alpha_{eu}$&$$&$$&$$&$$&$$&$$&$$&$$&$$&$$\\
${\cal Q}^1_\mu$&$ $&$$&$-\alpha_{qd}^{(1)}$&$$&$$&$$&$$&$$&$-\alpha_{lu}$&$$&$$&$$&$$&$$&$-\alpha_{qd}^{(1)}$\\
$ $&$ $&$$&$+\alpha_{qd}^{(8)}$&$$&$$&$$&$$&$$&$$&$$&$$&$$&$$&$$&$-\alpha_{qd}^{(8)}$\\
\crowcolor${\cal Q}^5_\mu$&$ $&$$&$-\alpha_{qu}^{(1)}$&$$&$$&$$&$$&$-\alpha_{ld}$&$-\alpha_{qe}$&$$&$$&$$&$$&$$&$-\alpha_{qu}^{(1)}$\\
\crowcolor$ $&$ $&$$&$+\alpha_{qu}^{(8)}$&$$&$$&$$&$$&$$&$$&$$&$$&$$&$$&$$&$-\alpha_{qu}^{(8)}$\\
${\cal X}_\mu$&$ $&$$&$$&$$&$-\alpha_{lq}^{(1)}$&$$&$$&$$&$$&$-\alpha_{lq}^{(1)}$&$$&$$&$$&$$&$$\\
$ $&$ $&$$&$$&$$&$-\alpha_{lq}^{(3)}$&$ $&$$&$$&$$&$+\alpha_{lq}^{(3)}$&$$&$$&$$&$$&$$\\
\crowcolor${\cal Y}^1_\mu$&$ $&$$&$-\alpha_{qd}^{(1)}$&$$&$$&$$&$$&$$&$$&$$&$$&$$&$$&$$&$-\alpha_{qd}^{(1)}$\\
\crowcolor$ $&$ $&$$&$-\alpha_{qd}^{(8)}$&$$&$$&$$&$$&$$&$$&$$&$$&$$&$$&$$&$+\alpha_{qd}^{(8)}$\\
${\cal Y}^5_\mu$&$ $&$$&$-\alpha_{qu}^{(1)}$&$$&$$&$$&$$&$$&$$&$$&$$&$$&$$&$$&$-\alpha_{qu}^{(1)}$\\
$ $&$ $&$$&$-\alpha_{qu}^{(8)}$&$$&$$&$$&$$&$$&$$&$$&$$&$$&$$&$$&$+\alpha_{qu}^{(8)}$\\
\cbottomrule
\end{tabular}
}
}
\vspace{1cm}
\rotatebox{90}{\parbox{23.25cm}{
\caption{
Four-fermion operators common to new scalar and vector fields. We only show operators that receive definite sign contributions from at least one particle, for $i=j$, $k=l$ ($i\not=k$). The symbols ``$+$'' (``$-$'') indicate that the contributions from scalars and vectors have the same (opposite) sign. We use ``$\pm$'' to indicate the absence of a definite sign in one of the sources. For the case of antisymmetric scalar couplings the relative sign is compared with the exact same index configuration as in Tables~\ref{Table:S0Table}-\ref{Table:PhiTable}.} 
\label{SV_4Fops}
}}
\end{center}
\end{table}
%

%-------------------------------- DOCUMENT: CONCLUSIONS----------------------------------%

\section{Conclusions}
\label{section_Conclusions}

The discovery at the LHC of a new particle of spin 0 has come hand in
hand with the direct observation of new interactions mediated by
scalar fields.\footnote{Indirect evidence of gauge-Higgs interactions
  was available before, in EWPD.} Among these, the Yukawa interactions
are quite unique in that they are not ruled by gauge invariance under
the SM gauge group, although of course they are compatible with
it. The exploration of this scalar sector is an important part of the
LHC physics program. The results at the LHC Run 1 have already
constrained significantly its structure. So far, all the measurements
are consistent with the minimal scalar sector of the SM: a Higgs
iso-doublet with a non-vanishing vev for its neutral component. But
the present uncertainties still allow for significant deviations in
the couplings of this doublet and for the presence of additional
scalar fields, related or not to electroweak symmetry breaking. To
comply with all the available data, such extra scalars must either
have small couplings or be significantly heavier than the Higgs
boson. The latter is the scenario we have considered in the second
part of this paper. 

We have followed a largely model-independent and unbiased approach,
with the minimal theoretical input of gauge-invariance. First, we have
classified into 19 irreducible representations all the possible scalar
fields that can interact linearly with the SM fields with
gauge-invariant renormalizable couplings. Their components with a
definite electric charge are the only scalar particles that can be
produced at colliders with unsuppressed couplings. We have written the
most general renormalizable interactions of the scalar multiplets
(except for parts of the scalar potential that cannot be tested in the
near future). Up to this point, all our results apply to either light
or heavy extra scalars. In a second step, we have assumed a hierarchy
of scales and have derived the dimension-six effective Lagrangian that
describes all the tree-level effects of the heavy scalars in
experiments where the probed energies are smaller than
their masses. We have shown that only the 19 scalar multiplets with
allowed linear interactions contribute to operators of dimension five
and six. Non-linear interactions of these fields also appear in the
effective Lagrangian to this order. The results are collected in
Appendix~B. Finally, we have used this effective Lagrangian to discuss
the observable effects of the new scalars and to derive bounds on
their couplings and masses. The strongest bounds come from flavor
observables. 
In order to avoid flavor constraints, here we have simply assumed that, in the flavor basis defined by
Eq. (\ref{SMLag}), there is only one non-vanishing entry in the Yukawa
couplings with the new scalar. We have then
studied the limits 
from a range of flavor-conserving observables: EWPD, LHC dilepton and
dijet searches and Higgs-mediated cross sections. 

Together with
Refs.~\cite{delAguila:2000rc,delAguila:2008pw,delAguila:2010mx}, this
paper provides a complete classification of all the particles with up to dimension-four linear couplings (in the electroweak symmetric phase) to the SM fields. Even if our emphasis in this paper has been on indirect effects, let us stress that this classification and the general interactions that are explicitly written in these references provide a useful basis for model-independent direct searches at large colliders (see e.g.~\cite{DelNobile:2009st,delAguila:2008cj,Atre:2008iu,Grojean:2011vu,deBlas:2012qp,Aguilar-Saavedra:2013qpa} for applications of this gauge-invariant formalism to direct searches).

% Adjust LaTeX output
\vspace{0.1cm}

On top of this, our results here complete the tree-level dictionary
between particles with general couplings of dimension $\leq 4$ and the effective operators that describe their low-energy effects. The dictionary entries for quarks, leptons and vector bosons can be found in Refs.~\cite{delAguila:2000rc,delAguila:2008pw,delAguila:2010mx}, respectively. We believe this correspondence can prove useful in combining the information from LHC searches of new particles with the existing precision constraints on their masses and couplings.

%------------------------------------ ACKNOWLEDGEMENTS ---------------------------------------%
\section*{Acknowledgements}

We thank L. Silvestrini and A. Weiler for useful discussions.
The work of JB has been supported by the European Research Council
under the European Union's Seventh Framework Programme
(FP~\!\!/~\!\!2007-2013)~\!/~\!ERC Grant Agreement n. 279972. The work
of MC, MPV and JS has been partially supported by the European
Commission through the contract PITN-GA-2012-316704 (HIGGSTOOLS), by
MINECO, under grant numbers FPA2010-17915 and FPA2013-47836-C3-2-P,
and by the Junta de Andaluc\'{\i}a grants FQM 101 and FQM 6552. 

%--------------------------------------------- APPENDICES -----------------------------------------------%
\newpage

\appendix

\clearpage

\section{Basis of dimension-six operators}
\label{app: L6_basis}

In this appendix, we introduce a complete set of gauge-invariant operators ${\cal O}_i$, which enter the general SM effective Lagrangian to dimension six:
\be
{\cal L}_\mt{Eff}^{(6)}={\cal L}_\SM+\frac{1}{\Lambda}{\cal L}_5+\frac{1}{\Lambda^2}{\cal L}_6,~\mbox{with}~~{\cal L}_d=\sum_i \alpha_i {\cal O}_i.\nonumber
\ee
We employ the basis in Tables~\ref{d45OpSM},~\ref{d6OpSM_4F}  and ~\ref{d6OpSM_Others}. Table~\ref{d45OpSM} defines
our notation for the effective operators renormalizing the SM interactions, and presents the unique dimension-five
interaction: the Weinberg operator, which gives Majorana masses to the SM neutrinos. Tables~\ref{d6OpSM_4F} and \ref{d6OpSM_Others} gather the dimension-six operators. In these tables, $T_A=\frac 12 \lambda_A$ and $f_{ABC}$, $A,B,C=1,\ldots,8$, are the $SU(3)_c$ generators and structure constants, with $\lambda_A$ the Gell-Mann matrices; $\epsilon_{ABC}$ ($\varepsilon_{abc}$) , $A,B,C=1,2,3$ ($a,b,c=1,2,3$) is the totally antisymmetric tensor in color (weak isospin) indices; $\sigma_a$, $a=1,2,3$ are the Pauli matrices; and $\widetilde{A}_{\mu\nu}=\frac12\varepsilon_{\mu\nu\rho\sigma}A^{\rho\sigma}$ is the Hodge-dual of the field strength $A_{\mu\nu}$. Finally, the superscript symbol ``$^\mt{T}$'' denotes transposition of the $SU(2)_L$ indices exclusively.\\ 

%Basis of dim 4-5
\begin{table}[h]
\begin{center}
\begin{tabular}{c c l c c l }
\ctoprule
&Operator& Notation& &Operator&Notation\\
\cmidrule{2-3}\cmidrule{5-6}
% Dimension 4
\multirow{3}{*}{\vtext{Dim 4~~~}}&$\left(\phi^{\dagger} \phi\right)^2$&${\cal O}_{\phi 4}$&&$\overline{l_L}~\!\phi~\!e_R$&${\cal O}_{y_e}$\\[0.1cm]
&&&&$\overline{q_L}~\!{\tilde \phi}~\!u_R$&${\cal O}_{y_u}$\\[0.1cm]
&&&&$\overline{q_L}~\!\phi~\!d_R$&${\cal O}_{y_d}$\\[0.1cm]
\cmrule
% Dimension 5
\multirow{2}{*}{\vtext{Dim 5}}&&&&&\\[-0.2cm]
&$\overline{l_L^c} \tilde{\phi}^* \tilde{\phi}^\dagger l_L$&${\cal O}_{5}$&&&\\[0.3cm]
\cbottomrule
\end{tabular}
\caption{Operators of dimension four and five.\label{d45OpSM}}
\end{center}
\end{table}
%

%Basis of dim 6
\begin{table}[thp]
\begin{center}
\begin{tabular}{c c l c c l }
\ctoprule
&Operator& Notation& &Operator&Notation\\
\cmidrule{2-3}\cmidrule{5-6}
%
% 4F: LLLL
\multirow{3}{*}{\vtext{LLLL}}&$\frac 1 2 \left(\overline{ l_L} \gamma_\mu l_L\right)\left(\overline{l_L} \gamma^\mu l_L\right)$&${\cal O}_{ll}^{(1)}$& &&\\
&$\frac 1 2 \left(\overline{q_L} \gamma_\mu q_L\right)\left(\overline{q_L} \gamma^\mu q_L\right)$&${\cal O}_{qq}^{(1)}$&&$\frac 1 2 \left(\overline{q_L} \gamma_\mu T_A q_L\right)\left(\overline{q_L} \gamma^\mu T_A q_L\right)$&${\cal O}_{qq}^{(8)}$\\
&$\left(\overline{l_L}\gamma_\mu l_L\right)\left(\overline{q_L} \gamma^\mu q_L\right)$&${\cal O}_{lq}^{(1)}$&&$\left(\overline{l_L} \gamma_\mu\sigma_a l_L\right)\left(\overline{q_L} \gamma^\mu\sigma_a q_L\right)$&${\cal O}_{lq}^{(3)}$\\[0.1cm]
\cmrule
% 4F: RRRR
\multirow{4}{*}{\vtext{RRRR}}&$\frac 1 2 \left(\overline{e_R} \gamma_\mu e_R\right)\left(\overline{e_R} \gamma^\mu e_R\right)$&${\cal O}_{ee}$&&\\
&$\frac 1 2 \left(\overline{u_R} \gamma_\mu u_R\right)\left(\overline{u_R} \gamma^\mu u_R\right)$&${\cal O}_{uu}^{(1)}$&&$\frac 1 2 \left(\overline{d_R} \gamma_\mu d_R\right)\left(\overline{d_R} \gamma^\mu d_R\right)$&${\cal O}_{dd}^{(1)}$\\
&$\left(\overline{u_R} \gamma_\mu u_R\right)\left(\overline{d_R} \gamma^\mu d_R\right)$&${\cal O}_{ud}^{(1)}$&&$\left(\overline{u_R} \gamma_\mu T_A u_R\right)\left(\overline{d_R} \gamma^\mu T_A d_R\right)$&${\cal O}_{ud}^{(8)}$\\
&$\left(\overline{e_R} \gamma_\mu e_R\right)\left(\overline{u_R} \gamma^\mu u_R\right)$&${\cal O}_{eu}$&&$\left(\overline{e_R} \gamma_\mu e_R\right)\left(\overline{d_R} \gamma^\mu d_R\right)$&${\cal O}_{ed}$\\[0.1cm]
\cmrule
% 4F: LLRR and LRRL
\multirow{5}{*}{\vtext{LLRR \& LRRL}}&$\left(\overline{l_L} \gamma_\mu l_L\right)\left(\overline{e_R} \gamma^\mu e_R\right)$&${\cal O}_{le}$&&$\left(\overline{q_L} \gamma_\mu q_L\right)\left(\overline{e_R} \gamma^\mu e_R\right)$&${\cal O}_{qe}$\\
&$\left(\overline{l_L} \gamma_\mu l_L\right)\left(\overline{u_R} \gamma^\mu u_R\right)$&${\cal O}_{lu}$&&$\left(\overline{l_L} \gamma_\mu l_L\right)\left(\overline{d_R} \gamma^\mu d_R\right)$&${\cal O}_{ld}$\\
&$\left(\overline{q_L} \gamma_\mu q_L\right)\left(\overline{u_R} \gamma^\mu u_R\right)$&${\cal O}_{qu}^{(1)}$&&$\left(\overline{q_L} \gamma_\mu T_A q_L\right)\left(\overline{u_R} \gamma^\mu T_A u_R\right)$&${\cal O}_{qu}^{(8)}$\\
&$\left(\overline{q_L} \gamma_\mu q_L\right)\left(\overline{d_R} \gamma^\mu d_R\right)$&${\cal O}_{qd}^{(1)}$&&$\left(\overline{q_L} \gamma_\mu T_A q_L\right)\left(\overline{d_R} \gamma^\mu T_A d_R\right)$&${\cal O}_{qd}^{(8)}$\\
&$\left(\overline{l_L} e_R\right)\left(\overline{d_R} q_L\right)$&${\cal O}_{ledq}$&& &\\[0.1cm]
\cmrule
%  4F: LRLR
\multirow{2}{*}{\vtext{LRLR}}&$\left(\overline{q_L}  u_R\right)i\sigma_2\left(\overline{q_L} d_R\right)^{\mt{T}}$&${\cal O}_{qud}^{(1)}$&&$\left(\overline{q_L} T_A u_R\right)i\sigma_2\left(\overline{q_L} T_A d_R\right)^{\mt{T}}$&${\cal O}_{qud}^{(8)}$\\
&$\left(\overline{l_L}  e_R\right)i\sigma_2\left(\overline{q_L} u_R\right)^{\mt{T}}$&${\cal O}_{lequ}$&&$\left(\overline{l_L} u_R\right)i\sigma_2\left(\overline{q_L} e_R\right)^{\mt{T}}$&${\cal O}_{luqe}$\\[0.1cm]
\cmrule
\multirow{3}{*}{\vtext{\BLOp~~~~}}&$\epsilon_{ABC}(\overline{l_L}i\sigma_2q_L^{c\ A})(\overline{d_R^{B}}u_R^{c\ C})$&${\cal O}_{lqdu}$&&$\epsilon_{ABC}(\overline{q_L^{A}}i\sigma_2 q_L^{c\ B})(\overline{e_R}u_R^{c\ C})$&
${\cal O}_{qqeu}$\\[0.1cm]
&$\epsilon_{ABC}(\overline{l_L}i\sigma_2 q_L^{c\ A})(\overline{q_L^B} i\sigma_2 q_L^{c\ C})$&${\cal O}_{lqqq}^{(1)}$&&$\epsilon_{ABC}(\overline{u_R^{A}} d_R^{c~\!B})(\overline{e_R} u_R^{c\ C})$&${\cal O}_{udeu}$\\[0.1cm]
&$\epsilon_{ABC}(\overline{l_L}\sigma_a i\sigma_2 q_L^{c\ A})(\overline{q_L^B} \sigma_a i\sigma_2 q_L^{c\ C})$&${\cal O}_{lqqq}^{(3)}$&&$$&$$\\[0.1cm]
\cbottomrule
\end{tabular}
\caption{Basis of dimension-six operators used in our analysis: four-fermion contact interactions. Flavor indices are omitted. \label{d6OpSM_4F}
}
\end{center}
\end{table}
%

%Basis of dim 6
\begin{table}[t]
\begin{center}
\begin{tabular}{c c l c c l }
\ctoprule
&Operator& Notation& &Operator&Notation\\
\cmidrule{2-3}\cmidrule{5-6}
% S
\vtext{S}&$\left(\phi^{\dagger} \phi\right)\!\Box\!\left(\phi^{\dagger} \phi\right)$&${\cal O}_{\phi \Box}$&&$\frac 1 3 \left(\phi^{\dagger} \phi\right)^3$&${\cal O}_{\phi }$\\[0.1cm]
\cmrule
%  SVF (Scalar-Vector-Fermion)
\multirow{5}{*}{\vtext{SVF}}&\tcolor{$\left(\phi^{\dagger}i D_\mu \phi\right)\left(\overline{l_L} \gamma^\mu l_L\right)$}&\tcolor{${\cal O}_{\phi l}^{(1)}$}&&\tcolor{$\left(\phi^{\dagger} \sigma_a iD_\mu \phi\right)\left(\overline{l_L} \gamma^\mu\sigma_a l_L\right)$}&\tcolor{${\cal O}_{\phi l}^{(3)}$}\\
&\tcolor{$\left(\phi^{\dagger}i D_\mu \phi\right)\left(\overline{e_R} \gamma^\mu e_R\right)$}&\tcolor{${\cal O}_{\phi e}^{(1)}$}&&$$&$$\\
&\tcolor{$\left(\phi^{\dagger}i D_\mu \phi\right)\left(\overline{q_L} \gamma^\mu q_L\right)$}&\tcolor{${\cal O}_{\phi q}^{(1)}$}&&\tcolor{$\left(\phi^{\dagger} \sigma_a iD_\mu \phi\right)\left(\overline{q_L} \gamma^\mu\sigma_a q_L\right)$}&\tcolor{${\cal O}_{\phi q}^{(3)}$}\\
&\tcolor{$\left(\phi^{\dagger}i D_\mu \phi\right)\left(\overline{u_R} \gamma^\mu u_R\right)$}&\tcolor{${\cal O}_{\phi u}^{(1)}$}&&\tcolor{$\left(\phi^{\dagger}i D_\mu \phi\right)\left(\overline{d_R} \gamma^\mu d_R\right)$}&\tcolor{${\cal O}_{\phi d}^{(1)}$}\\
&\tcolor{$\left(\phi^{T}i\sigma_2 iD_\mu \phi\right)\left(\overline{u_R} \gamma^\mu d_R\right)$}&\tcolor{${\cal O}_{\phi ud}$}&&$ $&$ $\\
\cmrule
%  STF (Scalar-Tensor-Fermion)
\multirow{4}{*}{\vtext{STF}}&\tcolor{$\left(\overline{l_L}\sigma^{\mu\nu}e_R\right)\phi~\! B_{\mu\nu}$}&\tcolor{${\cal O}_{eB}$}&&\tcolor{$\left(\overline{l_L}\sigma^{\mu\nu}e_R\right)\sigma^a\phi~\! W_{\mu\nu}^a$}&\tcolor{${\cal O}_{eW}$}\\
&\tcolor{$\left(\overline{q_L}\sigma^{\mu\nu}u_R\right)\tilde{\phi}~\! B_{\mu\nu}$}&\tcolor{${\cal O}_{uB}$}&&\tcolor{$\left(\overline{q_L}\sigma^{\mu\nu}u_R\right)\sigma^a\tilde{\phi}~\! W_{\mu\nu}^a$}&\tcolor{${\cal O}_{uW}$}\\
&\tcolor{$\left(\overline{q_L}\sigma^{\mu\nu}d_R\right)\phi~\! B_{\mu\nu}$}&\tcolor{${\cal O}_{dB}$}&&\tcolor{$\left(\overline{q_L}\sigma^{\mu\nu}d_R\right)\sigma^a\phi~\! W_{\mu\nu}^a$}&\tcolor{${\cal O}_{dW}$}\\
&\tcolor{$\left(\overline{q_L}\sigma^{\mu\nu}T_A u_R\right)\tilde{\phi}~\! G_{\mu\nu}^A$}&\tcolor{${\cal O}_{uG}$}&&\tcolor{$\left(\overline{q_L}\sigma^{\mu\nu}T_A d_R\right)\phi~\! G^A_{\mu\nu}$}&\tcolor{${\cal O}_{dG}$}\\[0.1cm]
\cmrule
%  SF
\multirow{2}{*}{\vtext{SF}}&$\left(\phi^{\dagger}\phi\right)\left(\overline{l_L}~\!\phi~\!e_R\right)$&${\cal O}_{e \phi }$&&$$&$$\\
&$\left(\phi^{\dagger}\phi\right)\left(\overline{q_L}~\!{\tilde \phi}~\!u_R\right)$&${\cal O}_{u \phi }$&&$\left(\phi^{\dagger}\phi\right)\left(\overline{q_L}~\!\phi~\!d_R\right)$&${\cal O}_{d \phi }$\\[0.1cm]
\cmrule
% Oblique
\multirow{5}{*}{\vtext{Oblique}}&$\left(\phi^{\dagger}D_\mu \phi\right)(\left(D^\mu \phi\right)^{\dagger}\phi)$&${\cal O}_{\phi D}$&&&\\
&\tcolor{$ \phi^\dagger \phi~B_{\mu\nu}B^{\mu\nu} $}&\tcolor{${\cal O}_{\phi{B}}$}&&\tcolor{$ \phi^\dagger \phi~\widetilde{B}_{\mu\nu}B^{\mu\nu} $}&\tcolor{${\cal O}_{\phi{\widetilde{B}}}$}\\
&\tcolor{$ \phi^\dagger \phi~W_{\mu\nu}^aW^{a~\!\mu\nu} $}&\tcolor{${\cal O}_{\phi{W}}$}&&\tcolor{$ \phi^\dagger \phi~\widetilde{W}_{\mu\nu}^aW^{a~\!\mu\nu} $}&\tcolor{${\cal O}_{\phi{\widetilde{W}}}$}\\
&\tcolor{$ \phi^\dagger\sigma_a \phi~W^a_{\mu\nu}B^{\mu\nu} $}&\tcolor{${\cal O}_{{WB}}$}&&\tcolor{$ \phi^\dagger\sigma_a \phi~\widetilde{W}^a_{\mu\nu}B^{\mu\nu} $}&\tcolor{${\cal O}_{\widetilde{W}B}$}\\
&\tcolor{$ \phi^\dagger \phi~G_{\mu\nu}^A G^{A~\!\mu\nu} $}&\tcolor{${\cal O}_{\phi{G}}$}&&\tcolor{$ \phi^\dagger \phi~\widetilde{G}_{\mu\nu}^A G^{A~\!\mu\nu} $}&\tcolor{${\cal O}_{\phi\widetilde{G}}$}\\[0.1cm]
\cmrule
%  TGC
\multirow{2}{*}{\vtext{Gauge}}&\tcolor{$\varepsilon_{abc}~\!W^{a~\!\nu}_\mu W^{b~\!\rho}_\nu W^{c~\!\mu}_\rho$}&\tcolor{${\cal O}_{W}$}&&\tcolor{$\varepsilon_{abc}~\!\widetilde{W}^{a~\!\nu}_\mu W^{b~\!\rho}_\nu W^{c~\!\mu}_\rho$}&\tcolor{${\cal O}_{\widetilde{W}}$}\\[0.1cm]
&\tcolor{$f_{ABC}~\!G^{A~\!\nu}_\mu G^{B~\!\rho}_\nu G^{C~\!\mu}_\rho$}&\tcolor{${\cal O}_{G}$}&&\tcolor{$f_{ABC}~\!\widetilde{G}^{A~\!\nu}_\mu G^{B~\!\rho}_\nu G^{C~\!\mu}_\rho$}&\tcolor{${\cal O}_{\widetilde{G}}$}\\[0.2cm]
\cbottomrule
\end{tabular}
\caption{Basis of dimension-six operators used in our analysis: operators other than four-fermion contact interactions. Flavor indices are omitted. Operators in grey color do not arise in the integration of heavy scalars at the tree level.
\label{d6OpSM_Others}
}
\end{center}
\end{table}

We use essentially the same basis as the one introduced in Ref.~\cite{Grzadkowski:2010es}. (See \cite{Buchmuller:1985jz,Contino:2013kra} for a related discussion of dimension-six physics in different operator bases.)  
The only differences (apart from changes in the names) are the use of different normalization factors in several operators
, and the trade of their operators $Q_{qq}^{(3)}=\left(\overline{q_L} \gamma_\mu \sigma_a q_L\right)\left(\overline{q_L} \gamma^\mu \sigma_a q_L\right)$ and $Q_{lequ}^{(3)}=\left(\overline{l_L} \sigma_{\mu\nu} e_R\right)i\sigma_2\left(\overline{q_L} \sigma^{\mu\nu} u_R\right)^{\mt{T}}$ by ${\cal O}_{qq}^{(8)}$ and ${\cal O}_{luqe}$, respectively, in our tables. Also, for consistency with previous works we write here the operators ${\cal O}_{\phi\psi}^{(1)}=\left(\phi^{\dagger}i D_\mu \phi\right)\left(\overline{\psi} \gamma^\mu \psi\right)$ and ${\cal O}_{\phi\psi}^{(3)}=\left(\phi^{\dagger}i \sigma_a D_\mu \phi\right)\left(\overline{\psi_L} \gamma^\mu \sigma_a \psi_L\right)$, instead of the hermitian interactions ${\cal Q}_{\phi\psi}^{(1)}=\left(\phi^{\dagger}i\lrD_\mu \phi\right)\left(\overline{\psi} \gamma^\mu \psi\right)$ and ${\cal Q}_{\phi\psi}^{(1)}=\left(\phi^{\dagger} i\lrDa_\mu \phi\right)\left(\overline{\psi_L} \gamma^\mu\sigma_a \psi_L\right)$ of Ref.~\cite{Grzadkowski:2010es}. Note that these last interactions are not generated in the integration of the new scalars, and are introduced here only for completeness.

Finally, for the purpose of comparing the results of the integration of new scalars with those obtained for new fermions and vector bosons in Refs.~\cite{delAguila:2000rc,delAguila:2008pw,delAguila:2010mx}, we provide below the necessary relations to translate the results in those references, which use the original basis of  \cite{Buchmuller:1985jz} and therefore contains redundant interactions, into our basis. Again, we use the notation ${\cal Q}_i$ to refer to the operator basis in other references, while we keep ${\cal O}_i$ for the operators presented in Tables~\ref{d6OpSM_4F} and \ref{d6OpSM_Others}.

In the sector of four-fermion interactions the following identities follow from the corresponding Fierz reorderings:

\be
\begin{array}{l c l }
\left({\cal Q}_{ll}^{(3)}\right)_{ijkl}&=&\frac 1 2 (\overline{l_L^i} \gamma_\mu \sigma_a l_L^j)(\overline{l_L^k} \gamma^\mu \sigma_a l_L^l)=2\left({\cal O}_{ll}^{(1)}\right)_{ilkj} - \left({\cal O}_{ll}^{(1)}\right)_{ijkl},\\
&&\\[-0.2cm]
\left({\cal Q}_{qq}^{(1,3)}\right)_{ijkl}&=&\frac 1 2 (\overline{q_L^i} \gamma_\mu \sigma_a q_L^j)(\overline{q_L^k} \gamma^\mu \sigma_a q_L^l)=-\left({\cal O}_{qq}^{(1)}\right)_{ijkl} +\frac 23 \left({\cal O}_{qq}^{(1)}\right)_{ilkj} +4 \left({\cal O}_{qq}^{(8)}\right)_{ilkj},\\
&&\\[-0.2cm]
\end{array}\nonumber
\ee
\be
\begin{array}{l c l }
\left({\cal Q}_{qq}^{(8,1)}\right)_{ijkl}&=&\frac 1 2 (\overline{q_L^i} \gamma_\mu \lambda_A q_L^j)(\overline{q_L^k} \gamma^\mu \lambda_A q_L^l)=4\left({\cal O}_{qq}^{(8)}\right)_{ijkl},\\
&&\\[-0.2cm]
\left({\cal Q}_{qq}^{(8,3)}\right)_{ijkl}&=&\frac 1 2 (\overline{q_L^i} \gamma_\mu \lambda_A \sigma_a q_L^j)(\overline{q_L^k} \gamma^\mu \lambda_A \sigma_a q_L^l)=\frac{32}{9}\left({\cal O}_{qq}^{(1)}\right)_{ilkj} -4 \left({\cal O}_{qq}^{(8)}\right)_{ijkl} - \frac 83\left({\cal O}_{qq}^{(8)}\right)_{ilkj},\\
&&\\[-0.2cm]
\left({\cal Q}_{ff}^{(8)}\right)_{ijkl}&=&\frac 1 2 (\overline{f_R^i} \gamma_\mu \lambda_A f_R^j)(\overline{f_R^k} \gamma^\mu \lambda_A f_R^l)=2\left({\cal O}_{ff}^{(1)}\right)_{ilkj} -\frac 23 \left({\cal O}_{ff}^{(1)}\right)_{ijkl},~~~~~(f=u,d)\\
&&\\[-0.2cm]
\left({\cal Q}_{Ff}\right)_{ijkl}&=&(\overline{F_L^i} f_R^j)(\overline{f_R^k} F_L^l)=-\frac 12\left({\cal O}_{Ff}\right)_{ilkj},~~~~~~~~~~~~~~~~~~~~~~~~~\!(Ff=le,lu,ld,qe)\\
&&\\[-0.2cm]
\left({\cal Q}_{qf}^{(1)}\right)_{ijkl}&=&(\overline{q_L^i} f_R^j)(\overline{f_R^k} q_L^l)=-\frac{1}{6}\left({\cal O}_{qf}^{(1)}\right)_{ilkj} - \left({\cal O}_{qf}^{(8)}\right)_{ilkj},\\
&&~~~~~~~~~~~~~~~~~~~~~~~~~~~~~~~~~~~~~~~~~~~~~~~~~~~~~~~~~~~~~~~~~~~~~~~~~~~~~(f=u,d)\\[-0.2cm]
\left({\cal Q}_{qf}^{(8)}\right)_{ijkl}&=&(\overline{q_L^i} \lambda_A f_R^j)(\overline{f_R^k}\lambda_A q_L^l)=-\frac{8}{9}\left({\cal O}_{qf}^{(1)}\right)_{ilkj} +\frac{2}{3}\left({\cal O}_{qf}^{(8)}\right)_{ilkj}.
\end{array}\nonumber
\ee
Finally, the operator ${\cal Q}_{qde}$ is labeled as ${\cal O}_{ledq}$ in Table~\ref{d6OpSM_4F}.

There are also some differences in the case of the bosonic operators. Firstly, the operators ${\cal Q}_{\phi 6}$ and ${\cal Q}_{\phi}^{(3)}$ correspond exactly to ${\cal O}_{\phi D}$ and ${\cal O}_{\phi}$, respectively. Secondly, using a perturbative field redefinition:
\be
\begin{split}
{\cal Q}_{\phi}^{(1)}&=\phi^\dagger \phi \left(D_\mu\phi\right)^\dagger D^\mu\phi=\frac 12 {\cal O}_{\phi\Box} - \mu_\phi^2 {\cal O}_{\phi 4} + 3 \lambda_\phi {\cal O}_{\phi} +\\
&+\frac 12\left(y^e_{ii}\left({\cal O}_{e\phi}\right)_{ii}+y^d_{ii}\left({\cal O}_{d\phi}\right)_{ii}+V_{ij}^\dagger y^u_{jj}\left({\cal O}_{u\phi}\right)_{ij}+\hc\right).
\end{split}\nonumber
\ee

\clearpage

\section{Operator coefficients in the effective Lagrangian}
\label{app: NS_OpCoeff}

In Tables~\ref{Table:S0Table}-\ref{Table:PhiTable} we present, for each new type of scalar, the contributions to the coefficients of the different dimension-six operators that result upon integration of one scalar multiplet at the tree level. 
Those contributions that arise only in the case where the theory contains several scalars at the same time (in the same or different representations) are given in Table~\ref{Table:MultiScalTable}. All these results are given in the basis in Appendix A. In some cases, this requires performing algebraic manipulations and field redefinitions on the operators that result directly from the integration.

Tables \ref{Table:S0Table}-\ref{Table:MultiScalTable} also contain the interactions in the high-energy Lagrangian, using the notation of Eq.~(\ref{LSMvarphi}). We only write here those interactions that are relevant for the computation of ${\cal L}_\mt{Eff}^{(6)}$. When gauge indices are explicitly shown, we use the following labeling for $SU(2)_L$ indices in the different representations: $\alpha,\beta=\frac 12, -\frac 12$ for $SU(2)_L$ doublets; $a,b,c=1,2,3$ for the components of $SU(2)_L$ triplets in Cartesian coordinates; and $I,J,K=\frac 32,\frac12,-\frac 12,-\frac 32$ for the components of the $SU(2)_L$ quadruplets. The matrices used to construct the different invariants are the following:

\begin{itemize}
{\item In constructing the triplets from two doublets we use the Pauli matrices
\bea
\sigma_{1}=\left(
\begin{array}{r r}
0&1\\
1&0
\end{array}
\right);
&
\sigma_{2}=\left(
\begin{array}{r r}
0&-i\\
i&0
\end{array}
\right);
&
\sigma_{3}=\left(
\begin{array}{r r}
1&0\\
0&-1
\end{array}
\right).\nonumber
\eea
}
{\item The isospin-1 product of two triplets is obtained through:
\be
f_{abc}=\frac{i}{\sqrt{2}}\varepsilon_{abc}.\nonumber
\ee
}
{\item Quadruplets are obtained from the product of an isospin-1 field and a doublet by means of 
\bea
C^{3/2}_{a\beta}=\frac{1}{\sqrt{2}}\left(
\begin{array}{r r}
1&0\\
-i&0\\
0&0
\end{array}
\right);
&
C^{1/2}_{a\beta}=\frac{1}{\sqrt{6}}\left(
\begin{array}{r r}
0&1\\
0&-i\\
-2&0
\end{array}
\right);\nonumber
\eea
\bea
C^{-1/2}_{a\beta}=-\frac{1}{\sqrt{6}}\left(
\begin{array}{r r}
1&0\\
i&0\\
0&2
\end{array}
\right);
&
C^{-3/2}_{a\beta}=-\frac{1}{\sqrt{2}}\left(
\begin{array}{r r}
0&1\\
0&i\\
0&0
\end{array}
\right).\nonumber
\eea
}
{\item The singlet product of two quadruplets is obtained through the $SU(2)$ product
\be
\epsilon_{IJ}=\frac 12 \left(
\begin{array}{c c c c }
0&0&0&1\\
0&0&-1&0\\
0&1&0&0\\
-1&0&0&0
\end{array}\right).\nonumber
\ee
}
\end{itemize}
Finally, for $SU(3)_c$ indices, we use the following notation for the symmetric product of colored fields:
$$\psi_1^{\left(A\right|}\ldots \psi_2^{\left|B\right)}\equiv \frac 12\left(\psi_1^{A}\ldots \psi_2^{B} + \psi_1^{B}\ldots \psi_2^{A}\right).$$

%-------------------------------------- Eff. Lag. Results --------------------------------------------

% table S0
\begin{table}[h]
\begin{center}
{\small
\begin{tabular}{c l c c l}
\ctoprule
\multicolumn{5}{c}{\Bfmath{{\cal S} \sim \left(1,1\right)_{0}}}\\
\multicolumn{5}{c}{}\\[-0.2cm]
\multicolumn{5}{c}{$\!V_{{\cal S}}=\kappa_{{\cal S}}~{\cal S} \phi^\dagger \phi + \lambda_{{\cal S}}~{\cal S}^2 \phi^\dagger \phi + \kappa_{{\cal S}^3}~{\cal S}^3 $}\\
\midrule
\multicolumn{5}{c}{{\bf Dimension-Four Operators}}\\
\multicolumn{5}{c}{$\alpha_{\phi 4}=\frac{\kappa_{\cal S}^2}{2M_{{\cal S}}^2}$}\\
$ $&$ $&&$ $&$ $\\[-0.3cm]
\midrule
\multicolumn{5}{c}{{\bf Scalar Operators}}\\
$ $&$ $&&$ $&$ $\\[-0.3cm]
\multicolumn{2}{l}{$\frac{\alpha_\phi}{\Lambda^2}=3\frac{\kappa_{\cal S}^2}{M_{\cal S}^2}\left(-\frac{\lambda_{\cal S}}{M_{\cal S}^2}+\frac{\kappa_{{\cal S}^3}\kappa_{\cal S}}{M_{\cal S}^4}\right)$}&&\multicolumn{2}{l}{$\frac{\alpha_{\phi \Box}}{\Lambda^2}=-\frac{\kappa_{\cal S}^2}{2M_{{\cal S}}^4}$}\\
$ $&$ $&&$ $&$ $\\[-0.3cm]
\cbottomrule
\end{tabular}
\caption{Operator coefficients arising from the integration of a ${\cal S}$ scalar field. See Table~\ref{Table:MultiScalTable} for collective contributions of several multiplets.}\label{Table:S0Table}
}
\end{center}
\end{table}
%

% table S1
\begin{table}[h]
\begin{center}
{\small
\begin{tabular}{c l c c l}
\ctoprule
\multicolumn{5}{c}{\Bfmath{{\cal S}_{1} \sim \left(1,1\right)_{1}}}\\
\multicolumn{5}{c}{}\\[-0.2cm]
\multicolumn{5}{c}{$\!J_{{\cal S}_{1}}=(y_{{\cal S}_{1}}^{l})_{ij}~\overline{l_L^i} i\sigma_2 l_L^{c~\!j}~~~\left((y_{{\cal S}_{1}}^{l})_{ij}=-(y_{{\cal S}_{1}}^{l})_{ji}\right) $}\\
\midrule
\multicolumn{5}{c}{{\bf Four-Fermion Operators: LLLL}}\\
$ $&$ $&&$ $&$ $\\[-0.3cm]
\multicolumn{5}{c}{$\frac{\left(\alpha_{ll}^{(1)}\right)_{ijkl}}{\Lambda^2}=2\frac{(y_{{\cal S}_{1}}^{l})_{ik}(y_{{\cal S}_{1}}^{l~\!\dagger})_{lj}}{M_{{\cal S}_{1}}^2}$}\\
$ $&$ $&&$ $&$ $\\[-0.3cm]
\cbottomrule
\end{tabular}
\caption{Operator coefficients arising from the integration of a ${\cal S}_{1}$ scalar field.}\label{Table:S1Table}
}
\end{center}
\end{table}
%

% table S2
\begin{table}[h]
\begin{center}
{\small
\begin{tabular}{c l c c l}
\ctoprule
\multicolumn{5}{c}{\Bfmath{{\cal S}_{2} \sim \left(1,1\right)_{2}}}\\
\multicolumn{5}{c}{}\\[-0.2cm]
\multicolumn{5}{c}{$\!J_{{\cal S}_{2}}=(y_{{\cal S}_{2}}^{e})_{ij}~\overline{e_R^i} e_R^{c~\!j}~~~\left((y_{{\cal S}_{2}}^{e})_{ij}=(y_{{\cal S}_{2}}^{e})_{ji}\right)$}\\
\midrule
\multicolumn{5}{c}{{\bf Four-Fermion Operators: RRRR}}\\
$ $&$ $&&$ $&$ $\\[-0.3cm]
\multicolumn{5}{c}{$\frac{\left(\alpha_{ee}\right)_{ijkl}}{\Lambda^2}=\frac{(y_{{\cal S}_{2}}^{e})_{ki}(y_{{\cal S}_{2}}^{e~\!\dagger})_{jl}}{M_{{\cal S}_{2}}^2}$}\\
$ $&$ $&&$ $&$ $\\[-0.3cm]
\cbottomrule
\end{tabular}
\caption{Operator coefficients arising from the integration of a ${\cal S}_{2}$ scalar field.}\label{Table:S2Table}
}
\end{center}
\end{table}

\clearpage

% table varphi
\begin{table}[t]
\begin{center}
{\small
\begin{tabular}{c l c c l}
\ctoprule
\multicolumn{5}{c}{\Bfmath{\varphi \sim \left(1,2\right)_{\frac 12}}}\\
\multicolumn{5}{c}{}\\[-0.4cm]
\multicolumn{5}{c}{$\!J_{\varphi}=(y_{{\varphi}}^{e})_{ij}~\overline{e_R^i}l_L^j + (y_{{\varphi}}^{d})_{ij}~\overline{d_R^i}q_L^j +(y_{{\varphi}}^{u})_{ij}~ i\sigma_2 \overline{q_L^i}^T u_R^j $}\\
\multicolumn{5}{c}{}\\[-0.4cm]
\multicolumn{5}{c}{$\!V_{\varphi}= \lambda_{{\varphi}}~(\varphi^\dagger \phi) (\phi^\dagger \phi)+\hc$}\\
\midrule
\multicolumn{5}{c}{{\bf Four-Fermion Operators:}}\\
$ $&$ $&&$ $&$ $\\[-0.5cm]
\multicolumn{1}{l}{$\bullet~${\bf LLRR}}&&$\phantom{~~Space~~}$&\multicolumn{1}{l}{}& \\
\multicolumn{5}{c}{}\\[-0.5cm]
$~\frac{\left(\alpha_{le}\right)_{ijkl}}{\Lambda^2}~=$&$\!\!\!\!-\frac{(y_{{\varphi}}^{e})_{kj}(y_{{\varphi}}^{e~\!\dagger})_{il}}{2M_{{\varphi}}^2}$&&$$&$$\\
$\frac{\left(\alpha_{qd}^{(1)}\right)_{ijkl}}{\Lambda^2}=$&$\!\!\!\!-\frac{(y_{{\varphi}}^{d})_{kj}(y_{{\varphi}}^{d~\!\dagger})_{il}}{6M_{{\varphi}}^2}$&&$\frac{\left(\alpha_{qd}^{(8)}\right)_{ijkl}}{\Lambda^2}=$&$\!\!\!\!6\frac{\left(\alpha_{qd}^{(1)}\right)_{ijkl}}{\Lambda^2}$\\
$\frac{\left(\alpha_{qu}^{(1)}\right)_{ijkl}}{\Lambda^2}=$&$\!\!\!\!-\frac{(y_{{\varphi}}^{u})_{il}(y_{{\varphi}}^{u~\!\dagger})_{kj}}{6M_{{\varphi}}^2}$&&$\frac{\left(\alpha_{qu}^{(8)}\right)_{ijkl}}{\Lambda^2}=$&$\!\!\!\!6\frac{\left(\alpha_{qu}^{(1)}\right)_{ijkl}}{\Lambda^2}$\\
\multicolumn{5}{c}{}\\[-0.4cm]
\multicolumn{1}{l}{$\bullet~${\bf LRRL}}&&$\phantom{~~Space~~}$&\multicolumn{1}{l}{$\bullet~${\bf LRLR}}& \\
\multicolumn{5}{c}{}\\[-0.5cm]
$\frac{\left(\alpha_{ledq}\right)_{ijkl}}{\Lambda^2}=$&$\!\!\!\!\frac{(y_{{\varphi}}^{d})_{kl}(y_{{\varphi}}^{e~\!\dagger})_{ij}}{M_{{\varphi}}^2}$&&$\frac{\left(\alpha_{lequ}\right)_{ijkl}}{\Lambda^2}=$&$\!\!\!\!\frac{(y_{{\varphi}}^{u})_{kl}(y_{{\varphi}}^{e~\!\dagger})_{ij}}{M_{{\varphi}}^2}$\\
$$&$$&&$\frac{\left(\alpha_{qud}^{(1)}\right)_{ijkl}}{\Lambda^2}=$&$\!\!\!\!-\frac{(y_{{\varphi}}^{u})_{ij}(y_{{\varphi}}^{d~\!\dagger})_{kl}}{M_{{\varphi}}^2}$\\
$ $&$ $&&$ $&$ $\\[-0.5cm]
\cmrule
\multicolumn{5}{c}{{\bf  Scalar-Fermion Operators}}\\
$\frac{\left(\alpha_{e\phi}\right)_{ij}}{\Lambda^2}=$&$\!\!\!\!\!\!\!\frac{\lambda_\varphi (y_{{\varphi}}^{e~\!\dagger})_{ij}}{M_{{\varphi}}^2}$&&$$&$$\\
$ $&$ $&&$ $&$ $\\[-0.5cm]
$\frac{\left(\alpha_{u\phi}\right)_{ij}}{\Lambda^2}=$&$\!\!\!\!\!\!\!-\frac{\lambda_\varphi^* (y_{{\varphi}}^{u})_{ij}}{M_{{\varphi}}^2}$&&$\frac{\left(\alpha_{d\phi}\right)_{ij}}{\Lambda^2}=$&$\!\!\!\!\!\!\!\frac{\lambda_\varphi (y_{{\varphi}}^{d~\!\dagger})_{ij}}{M_{{\varphi}}^2}$\\
$ $&$ $&&$ $&$ $\\[-0.5cm]
\cmrule
\multicolumn{5}{c}{{\bf  Scalar Operators}}\\
\multicolumn{5}{c}{ }\\[-0.45cm]
\multicolumn{5}{c}{$\frac{\alpha_{\phi}}{\Lambda^2}=3\frac{\left|\lambda_\varphi\right|^2}{M_{{\varphi}}^2}$}\\
$ $&$ $&&$ $&$ $\\[-0.475cm]
\cbottomrule
\end{tabular}
\caption{Operator coefficients arising from the integration of a $\varphi$ scalar field. See Table~\ref{Table:MultiScalTable} for collective contributions of several multiplets.
\label{Table:varphiTable}
}
}
\end{center}
\end{table}
%

% table Xi0
\begin{table}[b]
\begin{center}
{\small
\begin{tabular}{c c c c c}
\ctoprule
\multicolumn{5}{c}{\Bfmath{\Xi_{0} \sim \left(1,3\right)_{0}}}\\
\multicolumn{5}{c}{}\\[-0.3cm]
\multicolumn{5}{c}{$\!V_{\Xi_{0}}=\kappa_{{\Xi}_{0}}~\!\phi^\dagger \Xi_0^a \sigma_a \phi + \lambda_{{\Xi}_{0}}~\!\left(\Xi_0^a \Xi_0^a\right)\left(\phi^\dagger \phi\right)$}\\
\midrule
\multicolumn{5}{c}{{\bf Dimension-Four Operators}}\\
$ $&$ $&&$ $&$ $\\[-0.4cm]
$$&$$&$\alpha_{\phi 4}=\frac{\kappa_{\Xi_0}^2}{2M_{\Xi_{0}}^2}\left(1-4\frac{\mu_\phi^2}{M_{\Xi_{0}}^2}\right)$&$$&$$\\
$ $&$ $&&$ $&$ $\\[-0.5cm]
\cmrule
\multicolumn{5}{c}{{\bf  Scalar-Fermion Operators}}\\
$ $&$ $&&$ $&$ $\\[-0.4cm]
$\frac{\left(\alpha_{e\phi}\right)_{ij}}{\Lambda^2}=\frac{\kappa_{\Xi_0}^2y^e_{ii}}{M_{\Xi_{0}}^4}\delta_{ij}$&$$&$\frac{\left(\alpha_{u\phi}\right)_{ij}}{\Lambda^2}=\frac{\kappa_{\Xi_0}^2V_{ij}^\dagger y^u_{jj}}{M_{\Xi_{0}}^4}$&$$&$\frac{\left(\alpha_{d\phi}\right)_{ij}}{\Lambda^2}=\frac{\kappa_{\Xi_0}^2 y^d_{ii}}{M_{\Xi_{0}}^4}\delta_{ij}$\\
$ $&$ $&&$ $&$ $\\[-0.5cm]
\cmrule
\multicolumn{2}{c}{{\bf  Oblique Operators}}& &\multicolumn{2}{c}{{\bf Scalar Operators}}\\
\multicolumn{5}{c}{ }\\[-0.45cm]
$\frac{\alpha_{\phi D}}{\Lambda^2}=-2\frac{\kappa_{\Xi_0}^2}{M_{\Xi_{0}}^4}$&$$&&\multicolumn{2}{l}{$\!\!\!\frac{\alpha_{\phi \Box}}{\Lambda^2}=\frac{\kappa_{\Xi_0}^2}{2M_{\Xi_{0}}^4}$}\\
$ $&$ $&&$ $&$ $\\[-0.4cm]
$ $&$ $&&\multicolumn{2}{l}{$\frac{\alpha_{\phi}}{\Lambda^2}=-3\frac{ \kappa_{\Xi_0}^2}{M_{\Xi_0}^4}\left(\lambda_{\Xi_0}-2\lambda_\phi\right)$}\\
$ $&$ $&&$ $&$ $\\[-0.5cm]
\cbottomrule
\end{tabular}
\caption{Operator coefficients arising from the integration of a $\Xi_{0}$ scalar field. See Table~\ref{Table:MultiScalTable} for collective contributions of several multiplets.}\label{Table:Xi0Table}
}
\end{center}
\end{table}

\clearpage

% table Xi1
\begin{table}[b]
\begin{center}
{\small
\begin{tabular}{c c c c c}
\ctoprule
\multicolumn{5}{c}{\Bfmath{\Xi_{1} \sim \left(1,3\right)_{1}}}\\
\multicolumn{5}{c}{}\\[-0.2cm]
\multicolumn{5}{c}{$\!J_{\Xi_{1}}=(y_{{\Xi}_{1}}^{l})_{ij}~\overline{l_L^i} \sigma_a  i\sigma_2 l_L^{c~\!j}~~~\left((y_{{\Xi}_{1}}^{l})_{ij}=(y_{{\Xi}_{1}}^{l})_{ji}\right) $}\\
\multicolumn{5}{c}{}\\[-0.3cm]
\multicolumn{5}{c}{$\!V_{\Xi_{1}}=\left(\kappa_{{\Xi}_{1}}~\Xi_1^{a~\!\dagger} \left(\tilde{\phi}^\dagger \sigma_a \phi\right) +\hc\right) + \lambda_{{\Xi}_{1}}~\!\left(\Xi_1^{a~\!\dagger} \Xi_1^a\right)\left(\phi^\dagger \phi\right) + \tilde{\lambda}_{{\Xi}_{1}}~\!f_{abc} \left(\Xi_1^{a~\!\dagger} \Xi_1^b\right) \left(\phi^\dagger \sigma_c \phi\right)$}\\
\midrule
\multicolumn{5}{c}{{\bf Dimension Four and Five Operators}}\\
$ $&$ $&&$ $&$ $\\[-0.3cm]
\multicolumn{2}{l}{$\alpha_{\phi 4}=\frac{2\left|\kappa_{\Xi_1}\right|^2}{M_{\Xi_{1}}^2}\left(1-2\frac{\mu_\phi^2}{M_{\Xi_{1}}^2}\right)$}&&\multicolumn{2}{l}{$\frac{\left(\alpha_{5}\right)_{ij}}{\Lambda}=-2\frac{\kappa_{\Xi_1} \left(y_{\Xi_1}^{l~\!\dagger}\right)_{ij}}{M_{\Xi_1}^2}$}\\
$ $&$ $&&$ $&$ $\\[-0.3cm]
\cmrule
\multicolumn{5}{c}{{\bf Four-Fermion Operators: LLLL}}\\
$ $&$ $&&$ $&$ $\\[-0.3cm]
\multicolumn{5}{c}{$\frac{\left(\alpha_{ll}^{(1)}\right)_{ijkl}}{\Lambda^2}=2\frac{(y_{\Xi_1}^{l})_{ki}(y_{\Xi_1}^{l~\!\dagger})_{jl}}{M_{\Xi_1}^2}$}\\
$ $&$ $&&$ $&$ $\\[-0.3cm]
\cmrule
\multicolumn{5}{c}{{\bf  Scalar-Fermion Operators}}\\
$ $&$ $&&$ $&$ $\\[-0.3cm]
$\frac{\left(\alpha_{e\phi}\right)_{ij}}{\Lambda^2}=2\frac{\left|\kappa_{\Xi_1}\right|^2y^e_{ii}}{M_{\Xi_{1}}^4}\delta_{ij}$&&$\frac{\left(\alpha_{u\phi}\right)_{ij}}{\Lambda^2}=2\frac{\left|\kappa_{\Xi_1}\right|^2V_{ij}^\dagger y^u_{jj}}{M_{\Xi_{1}}^4}$&$$&$\frac{\left(\alpha_{d\phi}\right)_{ij}}{\Lambda^2}=2\frac{\left|\kappa_{\Xi_1}\right|^2y^d_{ii}}{M_{\Xi_{1}}^4}\delta_{ij}$\\
$ $&$ $&&$ $&$ $\\[-0.3cm]
\cmrule
\multicolumn{2}{c}{{\bf  Oblique Operators}}& &\multicolumn{2}{c}{{\bf Scalar Operators}}\\
\multicolumn{5}{c}{ }\\[-0.3cm]
\multicolumn{2}{l}{$\frac{\alpha_{\phi D}}{\Lambda^2}=4\frac{\left|\kappa_{\Xi_1}\right|^2}{M_{\Xi_{1}}^4}$}&&\multicolumn{2}{l}{$\!\!\!\frac{\alpha_{\phi\Box}}{\Lambda^2}=2\frac{\left|\kappa_{\Xi_1}\right|^2}{M_{\Xi_{1}}^4}$}\\
$ $&$ $&&$ $&$ $\\[-0.3cm]
$$&$$&&\multicolumn{2}{l}{$\frac{\alpha_{\phi}}{\Lambda^2}=-3\frac{\left|\kappa_{\Xi_1}\right|^2}{M_{\Xi_1}^4}\left(2\lambda_{\Xi_1}-\sqrt{2}\tilde{\lambda}_{\Xi_1}-4\lambda_\phi\right)$}\\
$ $&$ $&&$ $&$ $\\[-0.4cm]
\cbottomrule
\end{tabular}
\caption{Operator coefficients arising from the integration of a $\Xi_{1}$ scalar field. See Table~\ref{Table:MultiScalTable} for collective contributions of several multiplets.}\label{Table:Xi1Table}
}
\end{center}
\end{table}
%

% table Theta1
\begin{table}[t]
\begin{center}
{\small
\begin{tabular}{c l c c l}
\ctoprule
\multicolumn{5}{c}{\Bfmath{\Theta_{1} \sim \left(1,4\right)_{\frac 12}}}\\
\multicolumn{5}{c}{}\\[-0.3cm]
\multicolumn{5}{c}{$\!V_{\Theta_{1}}=\lambda_{{\Theta}_{1}}~\!\left(\phi^\dagger \sigma_a \phi\right) C^I_{a\beta}\tilde{\phi}_\beta \epsilon_{IJ}\Theta_1^{J} + \hc $}\\
\midrule
\multicolumn{5}{c}{{\bf Scalar Operators}}\\
$ $&$ $&&$ $&$ $\\[-0.3cm]
\multicolumn{5}{c}{$\frac{\alpha_{\phi}}{\Lambda^2}=\frac 12 \frac{\left|\lambda_{\Theta_1}\right|^2}{M_{\Theta_1}^2} $}\\
$ $&$ $&&$ $&$ $\\[-0.3cm]
\cbottomrule
\end{tabular}
\caption{Operator coefficients arising from the integration of a $\Theta_{1}$ scalar field. See Table~\ref{Table:MultiScalTable} for collective contributions of several multiplets.}\label{Table:Theta1Table}
}
\end{center}
\end{table}
%

% table Theta3
\begin{table}[h]
\begin{center}
{\small
\begin{tabular}{c l c c l}
\ctoprule
\multicolumn{5}{c}{\Bfmath{\Theta_{3} \sim \left(1,4\right)_{\frac 32}}}\\
\multicolumn{5}{c}{}\\[-0.3cm]
\multicolumn{5}{c}{$\!V_{\Theta_{3}}=\lambda_{{\Theta}_{3}}~\!\left(\phi^\dagger \sigma_a \tilde{\phi}\right)C^I_{a\beta} \tilde{\phi}_\beta \epsilon_{IJ}\Theta_3^{J}+\hc$}\\
\midrule
\multicolumn{5}{c}{{\bf Scalar Operators}}\\
$ $&$ $&&$ $&$ $\\[-0.3cm]
\multicolumn{5}{c}{$\frac{\alpha_{\phi}}{\Lambda^2}=\frac 32 \frac{\left|\lambda_{\Theta_3}\right|^2}{M_{\Theta_3}^2}$}\\
$ $&$ $&&$ $&$ $\\[-0.3cm]
\cbottomrule
\end{tabular}
\caption{Operator coefficients arising from the integration of a $\Theta_{3}$ scalar field. See Table~\ref{Table:MultiScalTable} for collective contributions of several multiplets.}\label{Table:Theta3Table}
}
\end{center}
\end{table}
%

% table omega1
\begin{table}[h]
\begin{center}
{\small
\begin{tabular}{c l c c l}
\ctoprule
\multicolumn{5}{c}{\Bfmath{{\omega}_{1} \sim \left(3,1\right)_{-\frac 13}}}\\
\multicolumn{5}{c}{}\\[-0.3cm]
\multicolumn{5}{c}{$J_{{\omega}_{1}}=(y_{{\omega}_{1}}^{ql})_{ij}~\!\overline{q_L^{c~\!i}}i\sigma_2 l_L^j + (y_{{\omega}_{1}}^{qq})_{ij}~\!\varepsilon_{ABC}~\!\overline{q_L^{i~\!B}} i\sigma_2 q_L^{c~\!j~\!C} + (y_{{\omega}_{1}}^{eu})_{ij}~\!\overline{e_R^{c~\!i}} u_R^j + (y_{{\omega}_{1}}^{du})_{ij}~\!\varepsilon_{ABC}~\!\overline{d_R^{i~\!B}} u_R^{c~\!j~\!C}$}\\
\multicolumn{5}{c}{}\\[-0.35cm]
\multicolumn{5}{c}{$\left((y_{{\omega}_{1}}^{qq})_{ij}=(y_{\omega_1}^{qq})_{ji}\right)$}\\
\midrule
\multicolumn{5}{c}{{\bf Four-Fermion Operators:}}\\
$ $&$ $&&$ $&$ $\\[-0.3cm]
\multicolumn{2}{l}{$\bullet~${\bf LLLL}}&\phantom{~~~~~~~~Space~~~~~~~~}&\multicolumn{2}{l}{$\bullet~${\bf RRRR}} \\
$\frac{\left(\alpha_{lq}^{(1)}\right)_{ijkl}}{\Lambda^2}=$&$\!\!\!\!\frac 14\frac{(y_{{\omega}_{1}}^{ql})_{lj}(y_{{\omega}_{1}}^{ql~\!\dagger})_{ik}}{M_{{\omega}_{1}}^2}$&&$\frac{\left(\alpha_{ud}^{(1)}\right)_{ijkl}}{\Lambda^2}=$&$\!\!\!\!\frac13\frac{(y_{{\omega}_{1}}^{du})_{ki}(y_{{\omega}_{1}}^{du~\!\dagger})_{jl}}{M_{{\omega}_{1}}^2}$\\
$ $&$ $&&$ $&$ $\\[-0.4cm]
$\frac{\left(\alpha_{lq}^{(3)}\right)_{ijkl}}{\Lambda^2}=$&$\!\!\!\!-\frac{\left(\alpha_{lq}^{(1)}\right)_{ijkl}}{\Lambda^2}$&&$\frac{\left(\alpha_{ud}^{(8)}\right)_{ijkl}}{\Lambda^2}=$&$\!\!\!\!-3\frac{\left(\alpha_{ud}^{(1)}\right)_{ijkl}}{\Lambda^2}$\\
$ $&$ $&&$ $&$ $\\[-0.4cm]
$\frac{\left(\alpha_{qq}^{(1)}\right)_{ijkl}}{\Lambda^2}=$&$\!\!\!\!\frac 43\frac{(y_{{\omega}_{1}}^{qq})_{ki}(y_{{\omega}_{1}}^{qq~\!\dagger})_{jl}}{M_{{\omega}_{1}}^2}$&&$~\frac{\left(\alpha_{eu}\right)_{ijkl}}{\Lambda^2}~\!=$&$\!\!\!\!\frac12\frac{(y_{{\omega}_{1}}^{eu})_{jl}(y_{{\omega}_{1}}^{eu~\!\dagger})_{ki}}{M_{{\omega}_{1}}^2}$\\
$ $&$ $&&$ $&$ $\\[-0.4cm]
$\frac{\left(\alpha_{qq}^{(8)}\right)_{ijkl}}{\Lambda^2}=$&$\!\!\!\!-3\frac{\left(\alpha_{qq}^{(1)}\right)_{ijkl}}{\Lambda^2}$&&$ $&$ $\\
$ $&$ $&&$ $&$ $\\[-0.4cm]
$ $&$ $&&$ $&$ $\\[-0.3cm]
\multicolumn{2}{l}{$\bullet~${\bf LRLR}}&\phantom{~~~~~~~~Space~~~~~~~~}&\multicolumn{2}{l}{{$\bullet$~\BLOpbf}} \\
$\frac{\left(\alpha_{qud}^{(1)}\right)_{ijkl}}{\Lambda^2}=$&$\!\!\!\!\frac 43\frac{(y_{{\omega}_{1}}^{qq})_{ki}(y_{{\omega}_{1}}^{du~\!\dagger})_{jl}}{M_{{\omega}_{1}}^2}$&&$\frac{\left(\alpha_{lqdu}\right)_{ijkl}}{\Lambda^2}=$&$\!\!\!\!-\frac{(y_{{\omega}_{1}}^{du})_{kl}(y_{{\omega}_{1}}^{ql~\!\dagger})_{ij}}{M_{{\omega}_{1}}^2}$\\
$ $&$ $&&$ $&$ $\\[-0.4cm]
$\frac{\left(\alpha_{qud}^{(8)}\right)_{ijkl}}{\Lambda^2}=$&$\!\!\!\!-3\frac{\left(\alpha_{qud}^{(1)}\right)_{ijkl}}{\Lambda^2}$&&$\frac{\left(\alpha_{qqeu}\right)_{ijkl}}{\Lambda^2}=$&$\!\!\!\!\frac{(y_{{\omega}_{1}}^{qq})_{ji}(y_{{\omega}_{1}}^{eu~\!\dagger})_{lk}}{M_{{\omega}_{1}}^2}$\\
$ $&$ $&&$ $&$ $\\[-0.4cm]
$\frac{\left(\alpha_{luqe}\right)_{ijkl}}{\Lambda^2}=$&$\!\!\!\!\frac{(y_{{\omega}_{1}}^{eu})_{lj}(y_{{\omega}_{1}}^{ql~\!\dagger})_{ik}}{M_{{\omega}_{1}}^2}$&&$\frac{\left(\alpha_{lqqq}^{(1)}\right)_{ijkl}}{\Lambda^2}=$&$\!\!\!\!-\frac{(y_{{\omega}_{1}}^{ql~\!\dagger})_{ij}(y_{{\omega}_{1}}^{qq})_{kl}}{M_{{\omega}_{1}}^2}$\\
$ $&$ $&&$ $&$ $\\[-0.4cm]
$\frac{\left(\alpha_{lequ}\right)_{ijkl}}{\Lambda^2}=$&$\!\!\!\!\frac{\left(\alpha_{luqe}\right)_{ilkj}}{\Lambda^2}$&&$\frac{\left(\alpha_{udeu}\right)_{ijkl}}{\Lambda^2}=$&$\!\!\!\!-\frac{(y_{{\omega}_{1}}^{eu~\!\dagger})_{lk}(y_{{\omega}_{1}}^{du})_{ji}}{M_{{\omega}_{1}}^2}$\\
$ $&$ $&&$ $&$ $\\[-0.4cm]
\cbottomrule
\end{tabular}
\caption{Operator coefficients arising from the integration of a ${\omega}_{1}$ scalar field.}\label{Table:om1Table}
}
\end{center}
\end{table}
%

% table omega2
\begin{table}[h]
\begin{center}
{\small
\begin{tabular}{c l c c l}
\ctoprule
\multicolumn{5}{c}{\Bfmath{{\omega}_{2} \sim \left(3,1\right)_{\frac 23}}}\\
\multicolumn{5}{c}{}\\[-0.2cm]
\multicolumn{5}{c}{$\!J_{{\omega}_{2}}=(y_{{\omega}_{2}}^{d})_{ij}~\varepsilon_{ABC}~\!\overline{d_R^{i~\!B}} d_R^{c~\!j~\!C}~~~\left((y_{{\omega}_{2}}^d)_{ij}=-(y_{\omega_2}^d)_{ji}\right)$}\\
\midrule
\multicolumn{5}{c}{{\bf Four-Fermion Operators: RRRR}}\\
$ $&$ $&&$ $&$ $\\[-0.3cm]
\multicolumn{5}{c}{$\frac{\left(\alpha_{dd}^{(1)}\right)_{ijkl}}{\Lambda^2}=2\frac{(y_{{\omega}_{2}}^{d})_{ki}(y_{{\omega}_{2}}^{d~\!\dagger})_{jl}}{M_{{\omega}_{2}}^2}$}\\
$ $&$ $&&$ $&$ $\\[-0.3cm]
\cbottomrule
\end{tabular}
\caption{Operator coefficients arising from the integration of a ${\omega}_{2}$ scalar field.
}\label{Table:om2Table}
}
\end{center}
\end{table}
%

% table omega4
\begin{table}[h]
\begin{center}
{\small
\begin{tabular}{c l c c l}
\ctoprule
\multicolumn{5}{c}{\Bfmath{{\omega}_{4} \sim \left(3,1\right)_{-\frac 43}}}\\
\multicolumn{5}{c}{}\\[-0.2cm]
\multicolumn{5}{c}{$\!J_{{\omega}_{4}}=(y_{{\omega}_{4}}^{ed})_{ij}~\overline{e_R^{c~\!i}} d_R^j + (y_{{\omega}_{4}}^{uu})_{ij}~\varepsilon_{ABC}~\!\overline{u_R^{i~\!B}} u_R^{c~\!j~\!C}~~~\left((y_{{\omega}_{4}}^{uu})_{ij}=-(y_{\omega_4}^{uu})_{ji}\right)$}\\
\midrule
\multicolumn{5}{c}{{\bf Four-Fermion Operators:}}\\
$ $&$ $&&$ $&$ $\\[-0.3cm]
\multicolumn{1}{l}{$\bullet~${\bf RRRR}}& &$\phantom{~~Space~~}$&\multicolumn{1}{l}{}& \\
$~\frac{\left(\alpha_{ed}\right)_{ijkl}}{\Lambda^2}~=$&$\!\!\!\!\frac{(y_{{\omega}_{4}}^{ed})_{jl}(y_{{\omega}_{4}}^{ed~\!\dagger})_{ki}}{2 M_{{\omega}_{4}}^2}$&&$\frac{\left(\alpha_{uu}^{(1)}\right)_{ijkl}}{\Lambda^2}=$&$\!\!\!\!2\frac{(y_{{\omega}_{4}}^{uu})_{ki}(y_{{\omega}_{4}}^{uu~\!\dagger})_{jl}}{M_{{\omega}_{4}}^2}$\\
$ $&$ $&&$ $&$ $\\[-0.3cm]
\multicolumn{1}{l}{$\bullet$~\BLOpbf}& &$\phantom{~~Space~~}$&\multicolumn{1}{l}{}& \\
$\frac{\left(\alpha_{udeu}\right)_{ijkl}}{\Lambda^2}=$&$\!\!\!\!2\frac{(y_{{\omega}_{4}}^{uu})_{il}(y_{{\omega}_{4}}^{ed~\!\dagger})_{jk}}{M_{{\omega}_{4}}^2}$&&$ $&$ $\\
$ $&$ $&&$ $&$ $\\[-0.3cm]
\cbottomrule
\end{tabular}
\caption{Operator coefficients arising from the integration of a ${\omega}_{4}$ scalar field.}\label{Table:om4Table}
}
\end{center}
\end{table}
%

% table Pi1
\begin{table}[h]
\begin{center}
{\small
\begin{tabular}{c l c c l}
\ctoprule
\multicolumn{5}{c}{\Bfmath{\Pi_{1} \sim \left(3,2\right)_{\frac 16}}}\\
\multicolumn{5}{c}{}\\[-0.2cm]
\multicolumn{5}{c}{$\!J_{\Pi_{1}}=(y_{\Pi_{1}}^{ld})_{ij}~~i\sigma_2\overline{l_L^i}^T d_R^j$}\\
\midrule
\multicolumn{5}{c}{{\bf Four-Fermion Operators: LLRR}}\\
$ $&$ $&&$ $&$ $\\[-0.3cm]
\multicolumn{5}{c}{$\frac{\left(\alpha_{ld}\right)_{ijkl}}{\Lambda^2}=-\frac{(y_{\Pi_{1}}^{ld})_{il}(y_{\Pi_{1}}^{ld~\!\dagger})_{kj}}{2 M_{\Pi_{1}}^2}$}\\
$ $&$ $&&$ $&$ $\\[-0.3cm]
\cbottomrule
\end{tabular}
\caption{Operator coefficients arising from the integration of a $\Pi_{1}$ scalar field. }\label{Table:Pi1Table}
}
\end{center}
\end{table}
%

% table Pi7
\begin{table}[h]
\begin{center}
{\small
\begin{tabular}{c l c c l}
\ctoprule
\multicolumn{5}{c}{\Bfmath{\Pi_{7} \sim \left(3,2\right)_{\frac 76}}}\\
\multicolumn{5}{c}{}\\[-0.2cm]
\multicolumn{5}{c}{$\!J_{\Pi_{7}}=(y_{\Pi_{7}}^{lu})_{ij}~i\sigma_2 \overline{l_L^i}^T u_R^j +(y_{\Pi_{7}}^{eq})_{ij}~ \overline{e_R^i} q_L^j$}\\
\midrule
\multicolumn{5}{c}{{\bf Four-Fermion Operators:}}\\
$ $&$ $&&$ $&$ $\\[-0.3cm]
\multicolumn{1}{l}{$\bullet~${\bf LLRR}}&&$\phantom{~~Space~~}$&\multicolumn{1}{l}{$\bullet~${\bf LRLR}}& \\
$\frac{\left(\alpha_{lu}\right)_{ijkl}}{\Lambda^2}=$&$\!\!\!\!-\frac{(y_{\Pi_{7}}^{lu})_{il}(y_{\Pi_{7}}^{lu~\!\dagger})_{kj}}{2M_{\Pi_{7}}^2}$&&$\frac{\left(\alpha_{luqe}\right)_{ijkl}}{\Lambda^2}=$&$\!\!\!\!-\frac{(y_{\Pi_{7}}^{lu})_{ij}(y_{\Pi_{7}}^{eq~\!\dagger})_{kl}}{M_{\Pi_{7}}^2}$\\
$\frac{\left(\alpha_{qe}\right)_{ijkl}}{\Lambda^2}=$&$\!\!\!\!-\frac{(y_{\Pi_{7}}^{eq})_{kj}(y_{\Pi_{7}}^{eq~\!\dagger})_{il}}{2M_{\Pi_{7}}^2}$&&$$&$ $\\
$ $&$ $&&$ $&$ $\\[-0.3cm]
\cbottomrule
\end{tabular}
\caption{Operator coefficients arising from the integration of a $\Pi_{7}$ scalar field.}\label{Table:Pi7Table}
}
\end{center}
\end{table}
%

% table zeta
\begin{table}[h]
\begin{center}
{\small
\begin{tabular}{c l c c l}
\ctoprule
\multicolumn{5}{c}{\Bfmath{\zeta \sim \left(3,3\right)_{-\frac 13}}}\\
\multicolumn{5}{c}{}\\[-0.2cm]
\multicolumn{5}{c}{$\!J_{\zeta}=(y_{\zeta}^{ql})_{ij}~\overline{q_L^{c~\!i}}i\sigma_2 \sigma_a l_L^j + (y_{\zeta}^{qq})_{ij}~\varepsilon_{ABC}~\!\overline{q_L^{i~\!B}}\sigma_a i\sigma_2 q_L^{c~\!j~\!C}~~~\left((y_{\zeta}^{qq})_{ij}=-(y_{\zeta}^{qq})_{ji}\right)$}\\
\midrule
\multicolumn{5}{c}{{\bf Four-Fermion Operators:}}\\
$ $&$ $&&$ $&$ $\\[-0.3cm]
\multicolumn{1}{l}{$\bullet~${\bf LLLL}}& &$\phantom{~~Space~~}$&\multicolumn{1}{l}{}& \\
$\frac{\left(\alpha_{lq}^{(1)}\right)_{ijkl}}{\Lambda^2}=$&$\!\!\!\!\frac 34\frac{(y_{\zeta}^{ql})_{lj}(y_{\zeta}^{ql~\!\dagger})_{ik}}{M_{\zeta}^2}$&&$\frac{\left(\alpha_{lq}^{(3)}\right)_{ijkl}}{\Lambda^2}=$&$\!\!\!\!\frac 13\frac{\left(\alpha_{lq}^{(1)}\right)_{ijkl}}{\Lambda^2}$\\
$\frac{\left(\alpha_{qq}^{(1)}\right)_{ijkl}}{\Lambda^2}=$&$\!\!\!\!\frac 43\frac{(y_{\zeta}^{qq})_{ki}(y_{\zeta}^{qq~\!\dagger})_{jl}}{M_{\zeta}^2}$&&$\frac{\left(\alpha_{qq}^{(8)}\right)_{ijkl}}{\Lambda^2}=$&$\!\!\!\!-3\frac{\left(\alpha_{qq}^{(1)}\right)_{ijkl}}{\Lambda^2}$\\
$ $&$ $&&$ $&$ $\\[-0.3cm]
\multicolumn{1}{l}{{$\bullet$~\BLOpbf}}& &$\phantom{~~Space~~}$&\multicolumn{1}{l}{}& \\
$\frac{\left(\alpha_{lqqq}^{(3)}\right)_{ijkl}}{\Lambda^2}=$&$\!\!\!\!-\frac{(y_{\zeta}^{qq})_{kl}(y_{\zeta}^{ql~\!\dagger})_{ij}}{M_{\zeta}^2}$&&$ $&$ $\\
$ $&$ $&&$ $&$ $\\[-0.3cm]
\cbottomrule
\end{tabular}
\caption{Operator coefficients arising from the integration of a $\zeta$ scalar field.}\label{Table:ZetaTable}
}
\end{center}
\end{table}
%

% table Omega1
\begin{table}[h]
\begin{center}
{\small
\begin{tabular}{c l c c l}
\ctoprule
\multicolumn{5}{c}{\Bfmath{\Omega_{1} \sim \left(6,1\right)_{\frac 13}}}\\
\multicolumn{5}{c}{}\\[-0.2cm]
\multicolumn{5}{c}{$\!J_{\Omega_{1}}=(y_{\Omega_1}^{ud})_{ij}~\overline{u_R^{c~\!i~\!\left(A\right|}} d_R^{j~\!\left|B\right)}+(y_{\Omega_1}^{qq})_{ij}~\overline{q_L^{c~\!i~\!\left(A\right|}}i\sigma_2 q_L^{j~\!\left|B\right)}~~~\left((y_{\Omega_1}^{qq})_{ij}=-(y_{\Omega_1}^{qq})_{ji}\right)$}\\
\midrule
\multicolumn{5}{c}{{\bf Four-Fermion Operators:}}\\
$ $&$ $&&$ $&$ $\\[-0.3cm]
\multicolumn{1}{l}{$\bullet~${\bf LLLL}}& &$\phantom{~~Space~~}$&\multicolumn{1}{l}{$\bullet~${\bf RRRR}}& \\
$\frac{\left(\alpha^{(1)}_{qq}\right)_{ijkl}}{\Lambda^2}=$&$\!\!\!\!\frac 23 \frac{(y_{\Omega_{1}}^{qq})_{jl}(y_{\Omega_{1}}^{qq~\!\dagger})_{ki}}{ M_{\Omega_{1}}^2}$&&$\frac{\left(\alpha^{(1)}_{ud}\right)_{ijkl}}{\Lambda^2}=$&$\!\!\!\!\frac 13 \frac{(y_{\Omega_{1}}^{ud})_{jl}(y_{\Omega_{1}}^{ud~\!\dagger})_{ki}}{ M_{\Omega_{1}}^2}$\\
$ $&$ $&&$ $&$ $\\[-0.3cm]
$\frac{\left(\alpha^{(8)}_{qq}\right)_{ijkl}}{\Lambda^2}=$&$\!\!\!\!\frac 32 \frac{\left(\alpha^{(1)}_{qq}\right)_{ijkl}}{\Lambda^2}$&&$\frac{\left(\alpha^{(8)}_{ud}\right)_{ijkl}}{\Lambda^2}=$&$\!\!\!\!\frac 32\frac{\left(\alpha^{(1)}_{ud}\right)_{ijkl}}{\Lambda^2}$\\
$ $&$ $&&$ $&$ $\\[-0.3cm]
$ $&$ $&&$ $&$ $\\[-0.3cm]
\multicolumn{1}{l}{$\bullet~${\bf LRLR}}&&$\phantom{~~Space~~}$&\multicolumn{1}{l}{ }& \\
$\frac{\left(\alpha_{qud}^{(1)}\right)_{ijkl}}{\Lambda^2}=$&$\!\!\!\!\frac 43 \frac{(y_{\Omega_{1}}^{ud})_{jl}(y_{\Omega_{1}}^{qq~\!\dagger})_{ik}}{ M_{\Omega_{1}}^2}$&&$\frac{\left(\alpha_{qud}^{(8)}\right)_{ijkl}}{\Lambda^2}=$&$\!\!\!\!\frac 32\frac{\left(\alpha_{qud}^{(1)}\right)_{ijkl}}{\Lambda^2}$\\
$ $&$ $&&$ $&$ $\\[-0.3cm]
\cbottomrule
\end{tabular}
\caption{Operator coefficients arising from the integration of a $\Omega_{1}$ scalar field.}\label{Table:Om1Table}
}
\end{center}
\end{table}
%

% table Omega2
\begin{table}[h]
\begin{center}
{\small
\begin{tabular}{c l c c l}
\ctoprule
\multicolumn{5}{c}{\Bfmath{\Omega_{2} \sim \left(6,1\right)_{-\frac 23}}}\\
\multicolumn{5}{c}{}\\[-0.2cm]
\multicolumn{5}{c}{$\!J_{\Omega_{2}}=(y_{\Omega_{2}}^d)_{ij}~\overline{d_R^{c~\!i~\!\left(A\right|}} d_R^{j~\!\left|B\right)}~~~\left((y_{\Omega_{2}}^d)_{ij}=(y_{\Omega_{2}}^d)_{ji}\right)$}\\
\midrule
\multicolumn{5}{c}{{\bf Four-Fermion Operators: RRRR}}\\
$ $&$ $&&$ $&$ $\\[-0.3cm]
\multicolumn{5}{c}{$\frac{\left(\alpha_{dd}^{(1)}\right)_{ijkl}}{\Lambda^2}=\frac{(y_{\Omega_{2}}^{d})_{jl}(y_{\Omega_{2}}^{d~\!\dagger})_{ki}}{ M_{\Omega_{2}}^2}$}\\
$ $&$ $&&$ $&$ $\\[-0.3cm]
\cbottomrule
\end{tabular}
\caption{Operator coefficients arising from the integration of a $\Omega_{2}$ scalar field.}\label{Table:Om2Table}
}
\end{center}
\end{table}
%

% table Omega4
\begin{table}[h]
\begin{center}
{\small
\begin{tabular}{c l c c l}
\ctoprule
\multicolumn{5}{c}{\Bfmath{\Omega_{4} \sim \left(6,1\right)_{\frac 43}}}\\
\multicolumn{5}{c}{}\\[-0.2cm]
\multicolumn{5}{c}{$\!J_{\Omega_{4}}=(y_{\Omega_4}^u)_{ij}~\overline{u_R^{c~\!i~\!\left(A\right|}} u_R^{j~\!\left|B\right)}~~~\left((y_{\Omega_4}^u)_{ij}=(y_{\Omega_4}^u)_{ji}\right)$}\\
\midrule
\multicolumn{5}{c}{{\bf Four-Fermion Operators: RRRR}}\\
$ $&$ $&&$ $&$ $\\[-0.3cm]
\multicolumn{5}{c}{$\frac{\left(\alpha_{uu}^{(1)}\right)_{ijkl}}{\Lambda^2}=\frac{(y_{\Omega_{4}}^{u})_{jl}(y_{\Omega_{4}}^{u~\!\dagger})_{ki}}{ M_{\Omega_{4}}^2}$}\\
$ $&$ $&&$ $&$ $\\[-0.3cm]
\cbottomrule
\end{tabular}
\caption{Operator coefficients arising from the integration of a $\Omega_{4}$ scalar field.}\label{Table:Om4Table}
}
\end{center}
\end{table}
%

% table Upsilon
\begin{table}[h]
\begin{center}
{\small
\begin{tabular}{c l c c l}
\ctoprule
\multicolumn{5}{c}{\Bfmath{\Upsilon \sim \left(6,3\right)_{\frac 13}}}\\
\multicolumn{5}{c}{}\\[-0.2cm]
\multicolumn{5}{c}{$\!J_{\Upsilon}=(y_{\Upsilon}^q)_{ij}~\overline{q_L^{c~\!i~\!\left(A\right|}}i\sigma_2  \sigma_a q_L^{j~\!\left|B\right)}~~~\left((y_{\Upsilon}^q)_{ij}=(y_{\Upsilon}^q)_{ji}\right)$}\\
\midrule
\multicolumn{5}{c}{{\bf Four-Fermion Operators: LLLL}}\\
$ $&$ $&&$ $&$ $\\[-0.3cm]
\multicolumn{2}{l}{$\frac{\left(\alpha_{qq}^{(1)}\right)_{ijkl}}{\Lambda^2}=\frac 43\frac{(y_{\Upsilon}^{q})_{lj}(y_{\Upsilon}^{q~\!\dagger})_{ik}}{M_{\Upsilon}^2}$}&&\multicolumn{2}{l}{$\frac{\left(\alpha_{qq}^{(8)}\right)_{ijkl}}{\Lambda^2}=\frac{3}{2}\frac{\left(\alpha_{qq}^{(1)}\right)_{ijkl}}{\Lambda^2}$}\\
$ $&$ $&&$ $&$ $\\[-0.3cm]
\cbottomrule
\end{tabular}
\caption{Operator coefficients arising from the integration of a $\Upsilon$ scalar field. }\label{Table:UpsTable}
}
\end{center}
\end{table}
%

% table Phi
\begin{table}[h]
\begin{center}
{\small
\begin{tabular}{c l c c l}
\ctoprule
\multicolumn{5}{c}{\Bfmath{\Phi \sim \left(8,2\right)_{\frac 12}}}\\
\multicolumn{5}{c}{}\\[-0.2cm]
\multicolumn{5}{c}{$\!J_{\Phi}=(y_\Phi^{qu})_{ij}~i\sigma_2 \overline{q_L^i}^T T_A u_R ^j+ (y_\Phi^{dq})_{ij}~\overline{d_R^i}T_A q_L^j$}\\
\midrule
\multicolumn{5}{c}{{\bf Four-Fermion Operators:}}\\
$ $&$ $&&$ $&$ $\\[-0.3cm]
\multicolumn{1}{l}{$\bullet~${\bf LLRR}}&&$\phantom{~~Space~~}$&\multicolumn{1}{l}{$\bullet~${\bf LRLR}}& \\
$\frac{\left(\alpha_{qu}^{(1)}\right)_{ijkl}}{\Lambda^2}=$&$\!\!\!\!-\frac 29\frac{(y_{\Phi}^{qu})_{il}(y_{\Phi}^{qu~\!\dagger})_{kj}}{M_{\Phi}^2}$&&$\frac{\left(\alpha_{qud}^{(8)}\right)_{ijkl}}{\Lambda^2}=$&$\!\!\!\!-\frac{(y_{\Phi}^{qu})_{ij}(y_{\Phi}^{dq~\!\dagger})_{kl}}{M_{\Phi}^2}$\\
$\frac{\left(\alpha_{qu}^{(8)}\right)_{ijkl}}{\Lambda^2}=$&$\!\!\!\!-\frac 34\frac{\left(\alpha_{qu}^{(1)}\right)_{ijkl}}{\Lambda^2}$&&$ $&$ $\\
$\frac{\left(\alpha_{qd}^{(1)}\right)_{ijkl}}{\Lambda^2}=$&$\!\!\!\!-\frac 29\frac{(y_{\Phi}^{dq})_{kj}(y_{\Phi}^{dq~\!\dagger})_{il}}{M_{\Phi}^2}$&&$ $&$ $\\
$\frac{\left(\alpha_{qd}^{(8)}\right)_{ijkl}}{\Lambda^2}=$&$\!\!\!\!-\frac 34\frac{\left(\alpha_{qd}^{(1)}\right)_{ijkl}}{\Lambda^2}$&&$ $&$ $\\
$ $&$ $&&$ $&$ $\\[-0.3cm]
\cbottomrule
\end{tabular}
\caption{Operator coefficients arising from the integration of a $\Phi$ scalar field.}\label{Table:PhiTable}
}
\end{center}
\end{table}
\clearpage

%Multiple scalars table
\begin{table}[h]
\begin{center}
{\small
\begin{tabular}{c l}
\ctoprule
\multicolumn{2}{c}{{\bf Mixed contributions from \Bfmath{\left\{{\cal S},~\varphi,~\Xi_0,~\Xi_1,~\Theta_1,~\Theta_3\right\}}}}\\
\multicolumn{2}{c}{}\\[-0.3cm]
\multicolumn{2}{l}{$\Delta {\cal L}_{\mt{int}}=
- \left(\varphi_i^\dagger J_{\varphi_i} + \Xi_{1i}^{a~\!\dagger} J_{\Xi_{1i}}^a +\hc \right)-\kappa_{\cal S}^i {\cal S}_i \phi^\dagger \phi -\kappa_{{\cal S}^3}^{ijk} {\cal S}_i {\cal S}_j {\cal S}_k - \kappa_{\Xi_0}^i \Xi_{0i}^a \phi^\dagger \sigma_a \phi$ }\\
\multicolumn{2}{l}{$  -\left(\kappa_{\Xi_1}^i \Xi_{1i}^{a~\! \dagger} \tilde{\phi}^\dagger \sigma_a \phi+\kappa_{{\cal S}\varphi}^{ij} {\cal S}_i \varphi^\dagger_{j}\phi+\hc\right)-\kappa_{{\cal S}\Xi_0}^{ijk} {\cal S}_i \Xi_{0j}^{a}\Xi_{0k}^{a}-\kappa_{{\cal S}\Xi_1}^{ijk} {\cal S}_i \Xi_{1j}^{a~\!\dagger}\Xi_{1k}^{a}$ }\\
\multicolumn{2}{l}{$-\kappa_{\Xi_0^3}^{ijk} f_{abc} \Xi_{0i}^a \Xi_{0j}^b \Xi_{0k}^c -\kappa_{\Xi_0\Xi_1}^{ijk} f_{abc}\Xi_{0i}^{a}\Xi_{1j}^{b~\!\dagger}\Xi_{1k}^{c}-\left(\kappa_{\Xi_0\varphi}^{ij} \Xi_{0i}^a \left(\varphi^\dagger_{j}\sigma_a\phi\right) + \kappa_{\Xi_1\varphi}^{ij} \Xi_{1i}^{a~\!\dagger} \left(\tilde{\varphi}^\dagger_{j}\sigma_a\phi\right) + \hc\right)$\!\!\! }\\
\multicolumn{2}{l}{$ -\left(\kappa_{\Xi_0 \Theta_1}^{ij}  \Xi_{0i}^{a} C_{a\beta}^I \tilde{\phi}_\beta \epsilon_{IJ}\Theta_{1j}^{J} + \kappa_{\Xi_1\Theta_1}^{ij}  \Xi_{1i}^{a~\!\dagger} C_{a\beta}^I \phi_\beta \epsilon_{IJ}\Theta_{1j}^{J} + \kappa_{\Xi_1\Theta_3}^{ij}  \Xi_{1i}^{a~\!\dagger} C_{a\beta}^I \tilde{\phi}_\beta \epsilon_{IJ} \Theta_{3j}^{J} + \hc\right)$ }\\
\multicolumn{2}{l}{$ -\lambda_{{\cal S}}^{ij} {\cal S}_i {\cal S}_j \left(\phi^\dagger \phi\right)- \left(\lambda_{\varphi}^i \left(\varphi_{i}^\dagger\phi\right) \left(\phi^\dagger \phi\right) +\hc\right) -\lambda_{\Xi_{0}}^{ij} \Xi_{0i}^a \Xi_{0j}^a  \left(\phi^\dagger \phi\right)-\tilde{\lambda}_{\Xi_{0}}^{ij} \Xi_{0i}^a \Xi_{0j}^b f_{abc} \left(\phi^\dagger \sigma_c\phi\right)$ }\\
\multicolumn{2}{l}{$-\lambda_{\Xi_{1}}^{ij} \Xi_{1i}^{a~\!\dagger} \Xi_{1j}^a \left(\phi^\dagger \phi\right) -\tilde{\lambda}_{\Xi_{1}}^{ij} f_{abc} \Xi_{1i}^{a~\!\dagger} \Xi_{1j}^b \left(\phi^\dagger \sigma_c\phi\right)- \lambda_{{\cal S}\Xi_{0}}^{ij} {\cal S}_i \Xi_{0j}^a \left(\phi^\dagger \sigma_a\phi\right)$}\\
\multicolumn{2}{l}{$-\left(\lambda_{{\cal S}\Xi_{1}}^{ij} {\cal S}_i \Xi_{1j}^{a~\!\dagger} \left(\tilde{\phi}^\dagger \sigma_a\phi \right)+ \lambda_{\Xi_1 \Xi_0}^{ij} f_{abc}\Xi_{1i}^{a~\!\dagger}\Xi_{0j}^{b}\left(\tilde{\phi}^\dagger\sigma_c\phi\right) +\hc\right)$ }\\
\multicolumn{2}{l}{$
-  \left(\lambda_{\Theta_1}^i    \left(\phi^\dagger \sigma_a\phi\right)C_{a\beta}^I\tilde{\phi}_\beta \epsilon_{IJ}\Theta_{1i}^J + \lambda_{\Theta_3}^i    \left(\phi^\dagger \sigma_a\tilde{\phi}\right)C_{a\beta}^I\tilde{\phi}_\beta \epsilon_{IJ}\Theta_{3i}^J +\hc\right) $ }\\
$ $&$ $\\[-0.3cm]
\midrule
\multicolumn{2}{c}{{\bf Scalar Operators}}\\
$ $&$ $\\[-0.3cm]
\multicolumn{2}{l}{$\frac{\alpha_{\phi}}{\Lambda^2}=\frac{3}{M_{\varphi_j}^2}\left|\lambda_\varphi^j-\frac{\kappa_{{\cal S} \varphi}^{ij} \kappa_{{\cal S}}^i}{M_{{\cal S}_{i}}^2}-\frac{\kappa_{\Xi_0 \varphi}^{ij} \kappa_{\Xi_0}^i}{M_{\Xi_{0i}}^2}-2\frac{(\kappa_{\Xi_1 \varphi}^{ij})^* \kappa_{\Xi_1}^i}{M_{\Xi_{1i}}^2}\right|^2 + \frac{1}{2 M_{\Theta_{1i}}^2}\left|\lambda_{\Theta_1}^i-\frac{\kappa_{\Xi_0}^j\kappa_{\Xi_0\Theta_1 }^{ji} }{M_{\Xi_{0j}}^2}-\frac{(\kappa_{\Xi_1}^j)^*\kappa_{ \Xi_1 \Theta_1}^{ji} }{M_{\Xi_{1j}}^2}\right|^2 $}\\
$ $&$+\frac{3}{2M_{\Theta_{3i}}^2}\left|\lambda^{i}_{\Theta_3}-\frac{(\kappa_{\Xi_1}^j)^*\kappa_{\Xi_1\Theta_3 }^{ji} }{M_{\Xi_{1j}}^2}\right|^2 - \frac{3\kappa_{\cal S}^i}{M_{{\cal S}_i}^2}\left(\frac{ \lambda_{\cal S}^{ij}\kappa_{\cal S}^j}{M_{{\cal S}_j}^2}+\frac{ \lambda_{{\cal S}\Xi_0}^{ij}\kappa_{\Xi_0}^j}{M_{\Xi_{0j}}^2}+4\frac{\mt{Re}\left\{ \lambda_{{\cal S}\Xi_1}^{ij}(\kappa_{\Xi_1}^j)^*\right\}}{M_{\Xi_{1j}}^2}\right)-3\frac{\lambda_{\Xi_0}^{ij}\kappa_{\Xi_0}^i\kappa_{\Xi_0}^j}{M_{\Xi_{0i}}^2 M_{\Xi_{0j}}^2}$\\
$ $&$-3\frac{(\kappa_{\Xi_1}^i)^*\kappa_{\Xi_1}^j}{M_{\Xi_{1i}}^2 M_{\Xi_{1j}}^2}\left(2\lambda_{\Xi_1}^{ij}-\sqrt{2}\tilde{\lambda}_{\Xi_1}^{ij}\right)-6\sqrt{2}\frac{\mt{Re}\left\{\lambda_{\Xi_1\Xi_0}^{ij}(\kappa_{\Xi_1}^i)^* \kappa_{\Xi_0}^{j}\right\}}{M_{\Xi_{1i}}^2 M_{\Xi_{0j}}^2}-3\sqrt{2}\frac{\kappa_{\Xi_0\Xi_1}^{ijk}\kappa_{\Xi_0}^i(\kappa_{\Xi_1}^j)^* \kappa_{\Xi_1}^{k}}{M_{\Xi_{0i}}^2 M_{\Xi_{1j}}^2 M_{\Xi_{1k}}^2 }$\\
$ $&$+3\frac{\kappa_{\cal S}^i}{M_{{\cal S}_i}^2}\left(\frac{\kappa_{{\cal S}}^{ijk}\kappa_{{\cal S}}^j\kappa_{{\cal S}}^{k}}{M_{{\cal S}_j}^2 M_{{\cal S}_k}^2} +\frac{\kappa_{{\cal S} \Xi_0}^{ijk}\kappa_{\Xi_0}^j \kappa_{\Xi_0}^{k}}{M_{\Xi_{0j}}^2 M_{\Xi_{0k}}^2} +2\frac{\kappa_{{\cal S}\Xi_1}^{ijk}(\kappa_{\Xi_1}^j)^* \kappa_{\Xi_1}^{k}}{M_{\Xi_{1j}}^2 M_{\Xi_{1k}}^2}\right) $\\
$ $&$$\\[-0.3cm]
\midrule
\multicolumn{2}{c}{{\bf Scalar-Fermion Operators}}\\
$ $&$ $\\[-0.3cm]
\multicolumn{2}{l}{$\frac{\left(\alpha_{e\phi}\right)_{ij}}{\Lambda^2}=\frac{1}{M_{\varphi_j}^2}\left(\lambda_{\varphi}^j-\frac{\kappa_{{\cal S}}^i \kappa_{{\cal S}\varphi}^{ij}}{M_{{\cal S}_{i}}^2}-\frac{\kappa_{\Xi_0}^i \kappa_{\Xi_0\varphi}^{ij}}{M_{\Xi_{0i}}^2}-2\frac{\kappa_{\Xi_1}^i\left(\kappa_{\Xi_1\varphi}^{ij}\right)^*}{M_{\Xi_{1i}}^2}\right) \left(y_\varphi^{e~\!\dagger}\right)_{ij}$}\\
$ $&$ $\\[-0.3cm]
\multicolumn{2}{l}{$\frac{\left(\alpha_{d\phi}\right)_{ij}}{\Lambda^2}=\frac{1}{M_{\varphi_j}^2}\left(\lambda_{\varphi}^j-\frac{\kappa_{{\cal S}}^i \kappa_{{\cal S}\varphi}^{ij}}{M_{{\cal S}_{i}}^2}-\frac{\kappa_{\Xi_0}^i \kappa_{\Xi_0\varphi}^{ij}}{M_{\Xi_{0i}}^2}-2\frac{\kappa_{\Xi_1}^i\left(\kappa_{\Xi_1\varphi}^{ij}\right)^*}{M_{\Xi_{1i}}^2}\right) \left(y_\varphi^{d~\!\dagger}\right)_{ij}$}\\
$ $&$ $\\[-0.3cm]
\multicolumn{2}{l}{$\frac{\left(\alpha_{u\phi}\right)_{ij}}{\Lambda^2}=-\frac{1}{M_{\varphi_j}^2}\left((\lambda_{\varphi}^j)^*-\frac{\kappa_{{\cal S}}^i\left(\kappa_{{\cal S}\varphi}^{ij}\right)^*}{M_{{\cal S}_{i}}^2}-\frac{\kappa_{\Xi_0}^i\left(\kappa_{\Xi_0\varphi}^{ij}\right)^*}{M_{\Xi_{0i}}^2}-2\frac{\left(\kappa_{\Xi_1}^i\right)^*\kappa_{\Xi_1\varphi}^{ij}}{M_{\Xi_{1i}}^2}\right) \left(y_\varphi^u\right)_{ij}$}\\
$ $&$ $\\[-0.3cm]
\cbottomrule
\end{tabular}
\caption{Collective contributions from several scalars to the dimension six effective Lagrangian. Only those operators that receive contributions from interactions involving more than one scalar at a time are shown. The full contribution to the operator coefficient is presented in that case. For the other operators the contributions can be read by adding the corresponding pieces from the tables obtained integrating one scalar at a time. Note that, due to the antisymmetric properties of the interaction, the couplings $\kappa_{\Xi_0^3}^{ijk}$ and $\tilde{\lambda}_{\Xi_0}^{ij}$ do not contribute to ${\cal L}_6$.}\label{Table:MultiScalTable}
}
\end{center}
\end{table}

\clearpage

\newpage
%------------------------------------------- REFERENCES -------------------------------------------%
% Refs. Page 1: Adjust LaTeX output
~\vspace{-0.2cm}
%

%-------------------------------------------------- END --------------------------------------------------%

\end{document}